\documentclass[usenatbib,fleqn]{mn2e}
\usepackage{amsmath}
\usepackage{graphicx}
\usepackage{subfigure}
\usepackage{float}

\title[The Mass-Metallicity Relation at High Redshift]{The Origin and Evolution of the Mass-Metallicity Relation at High Redshift using GalICS}

\author[J.Sakstein et al.]{Jeremy Sakstein$^{1,2}$, Antonio Pipino$^{1,3,4}$,
Julien E.G. Devriendt$^1$ and Roberto Maiolino$^5$\\\\
$^1$Department of Astrophysics, University of Oxford, Denys Wilkinson Building,
    Keble Road, Oxford, OX1 3RH, U.K.\\
$^2$Department of Applied Mathematics and Theoretical Physics, University of Cambridge,\\ \hspace{1.3mm}Centre for Mathematical Sciences, Wilberforce Road, Cambridge CB3 0WA, U.K.\\
$^3$Dip. Fisica, sez. Astronomia, Universita di Trieste, via G.B. Tiepolo 11, 34100 Trieste, Italy.\\
$^4$Dept. of Physics and Astronomy, University of California, Los Angeles, 430 Portola Plaza, Los Angeles 90095, U.S.A.\\
$^5$INAF - Osservatorio Astronomico di Roma, via di Frascati 33,
00040 Monte Porzio Catone, Italy.}
\date{Accepted,
      Received }

\begin{document}
\maketitle

\begin{abstract}
\noindent The \textsc{G}al\textsc{ICS} (Galaxies in Cosmological Simulations) semi-analytical model of hierarchical galaxy formation is used to investigate the effects of different galactic properties, including star formation rate (SFR) and outflows, on the shape of the mass metallicity relation and to predict the relation for galaxies at redshift $z=2.27$ and $z=3.54$. Our version of \textsc{G}al\textsc{ICS} has the chemical evolution implemented in great detail and is less heavily reliant on approximations such as instantaneous recycling.  We vary the model parameters controlling both the efficiency and redshift dependence of the SFR as well as the efficiency of supernova feedback. We find that the factors controlling the SFR influence the relation significantly at all redshifts and require a strong redshift dependence, proportional to $1+z$, in order to reproduce the observed relation at the low mass end. Indeed, at any redshift, the predicted relation flattens out at the high mass end resulting in a poorer agreement with observations in this regime. We also find that variation of the parameters associated with outflows has a minimal effect on the relation at high redshift but does serve to alter its shape in the more recent past. We thus conclude that the relation is one between SFR and mass and that outflows are only important in shaping the relation at late times. When the relation is stratified by SFR it is apparent that the predicted galaxies with increasing stellar masses have higher SFRs, supporting the view that galaxy downsizing is the origin of the relation. Attempting to reproduce the observed relation, we vary the parameters controlling the efficiency of star formation and its redshift dependence and compare the predicted relations with \citet{erb06} at $z=2.27$ and \citet{amaze} at $z=3.54$ in order to find the best-fitting parameters. We succeed in fitting the relation at $z=3.54$ reasonably well, however we fail at $z=2.27$, our relation lying on average below the observed one at the one standard deviation level. We do, however, predict the observed evolution between $z=3.54$ and $z=0$. Finally, we discuss the reasons for the above failure and the flattening at high masses, with regards to both the comparability of our predictions with observations and the possible lack of underlying physics. Several of these problems are common to many semi-analytic/hybrid models and so we discuss possible improvements and set the stage for future work by considering how the predictions and physics in these models can be made more robust in light of our results.

\end{abstract}

\begin{keywords}
methods: N-body simulations -- galaxies: evolution -- galaxies:
haloes -- galaxies: formation
\end{keywords}

\section{Introduction}\label{sec:intro}

The existence of a relation between luminosity (or mass) and
metallicity in irregular and blue compact galaxies was first
proposed by \citet{leq79} and later confirmed by \citet{skillman89}.
\citet{garnett87} extended the relation to spiral galaxies. Recently
\citet{trem04} examined the relation at $z\!\approx\!0$ using
$\sim$53000 local star-forming galaxies in the SDSS and found that
$12+\log(O/H)$ increases steeply for
$10^{8.5}<M_\star/M_{\odot}<10^{10.5}$ but flattens at larger
masses.\newline\newline Several explanations for such a relation
have been put forward. For instance, \cite{trem04} suggest that
galaxies form similar fractions of stars independent of their mass.
Less massive galaxies are simply less efficient at retaining gas due
to their shallower potential wells and lose newly produced metals to
galactic outflows. Another possibility \citep{amaze} is that more
massive galaxies had higher specific star formation rates (SFR per
unit stellar mass) or a higher star formation efficiency (SFR per
unit gas mass) in the past so form a larger fraction of stars in the
same Hubble time. In this case the higher metallicity is due to the
conversion of a larger amount of primordial gas into C, N and O.
This scenario is known as galaxy down-sizing for which there is a
plethora of observational evidence (e.g \citet{Gonzallez07}). An
alternate explanation \citep{kop07} is a variation in the initial
mass function (IMF) in different star forming environments. It is
not known at present which process is responsible for the
ubiquitously observed relation however it is likely that all three
contribute to some extent. Interestingly, a similar relation holds
for the stars of gas-poor galaxies \citep{faber73,brod91}.\newline

A lot of effort has been made to understand whether an underlying
common physical origin for the above mentioned relation exists, as
well as to determine the highest redshift at which the
mass-metallicity relation holds. \citet{erb06} have measured the relation
at $z\approx2.27$ using the [NII]/H$\alpha$ ratio in stacked spectra
for a sample of 87 rest-frame ultraviolet-selected star-forming
galaxies. Recently, \citet{amaze} have used deep near-IR spectroscopy
of H$\beta$ and [OIII]5007 shifted into the \textit{K} band as well as
[OII]3727 and [NeIII]2870 shifted into the \textit{H} band to measure the
relation for nine star forming galaxies at $z\sim3.54$. More recently, \citet{Mann} have used near-IR spectroscopy of the optical lines [OII]$\lambda$3727, H$\beta$, and [OIII]$\lambda$5007 for a sample of 10 Lyman-break galaxies (LGBs) at $z\sim3$ to derive their SFR, metallicity, gas fraction and effective yield. Using optical, near-
IR and Spitzer/IRAC photometry, they have measured the stellar mass of each galaxy in order to
guarantee a robust estimate of the mass-metallicity relation. Finally, \citet{mann_new} introduce the SFR-mass-metallicity relation
and show that it does not evolve with redshift up to $z\sim2$. They argue that the apparent evolution of the mass-metallicity
relation inferred by past works at redshifts below $z \sim 2$ is only due to selection effects, where independent surveys sample different areas of the SFR-mass-metallicity ``surface'' at different redshifts (in particular those with a higher SFR at higher redshifts). A strong evolution still occurs for $z>2$.
\newline

Previous attempts at modelling hierarchical galaxy evolution through
semi-analytic models and $N$-body simulations have shown that a
mass-metallicity relation is predicted also in the hierarchical
scenario \citep{derossi}. In one such simulation \citet{derossi}
predicted the relation over the redshift range $0\le z\le3$; however the
predicted metallicities are always larger than the observed ones.
The authors attribute this to the lack of supernova feedback in the model and thus the
lack of outflows from the galaxies.
\citet{del04} predicts a relation at $z\approx0$ for three
 models characterized by different feedback processes. It was found that all three models predict galaxies whose average
metallicities lie within one standard deviation of the median
mass-metallicity relation observed by \citet{trem04}, at variance with the results of \citet{findave08}, who found that the predicted relation
depends strongly on the galactic outflow model used.
They also found that a lack of outflows lead to galaxies that were too enriched in metals, consistent with the results of \citet{derossi}.
According to \citet{findave08} only a model where the wind was momentum
driven could reproduce the observed relation \citep{erb06}. In this
simulation the relation was reproduced reasonably well within experimental
uncertainties and could match the observed relation in slope, amplitude and scatter.
By incorporating mass dependent galactic winds into
the parallel tree-SPH code \textsc{GADGET-2} \citep{gadget}, \citet{koba}
predicted a relation at $z\approx 2$ that is consistent with the
observations of \citet{erb06} for massive galaxies only.
Their predictions exhibit
a significant scatter and their model suffers from the fact that it does not predict the
termination of star formation in massive galaxies at late times.
Finally, using $N$-body/hydrodynamical simulations \citet{Mouchine} were able to
reproduce the observed relations over the range $0\le z\le1.2$ with
minimal scatter although at higher redshifts the simulated relation
predicts galaxies with higher metallicities than are observed
\citep{erb06}.\newline

As noted in \citet{amaze}, where we address the reader
for a more extended discussion of the single cases, many of the above mentioned attempts have failed to reproduce the observed
relation at $z\sim3$. \citet{calura} have recently been successful in predicting the relation over the range $0.07\le z\le3.54$ using a chemical evolution model that predicts the relation separately for galaxies of different morphological type. Rather than collectively fitting the relation they have found that the relation at low redshifts is best fit by considering only spiral and irregular galaxies whilst at intermediate redshifts ($z\sim2$) the relation is best fit by a mixture of proto-spirals and proto-ellipticals. At $z=3.54$ they predict that the relation is best fit by proto-ellipticals alone. In this work we
attempt to reproduce and explain the observed relation at $z=2.27$
and $z=3.54$ by testing aspects of the galactic outflows (in common with many of the above models) as well
as other methods including different IMFs and a redshift dependent
SFR. In order to do this, we make use of an up-to-date version of
{\sc G}al{\sc ICS} in which a detailed treatment of the chemical evolution has
been implemented; details of which can be found in \citet{galics2}. This model - which tracks the evolution of H, He, O and Fe - has the chemical evolution implemented in great detail. It does not rely on the instantaneous recycling approximation but instead uses a self-consistent prescription for both type Ia and Type II supernovae ejecta, which includes the effects of finite stellar lifetimes. We consider this aspect the main innovation of our approach.  \newline

We assume a flat (critical density) lambda cold dark matter
cosmology ($\Lambda$CDM) with cosmological parameters taken from the
WMAP three year results (WMAP3) \citep{wmap3}. These are
$\Omega_0=0.24$, $\Omega_{\Lambda}=0.76$, $h=0.73$ and
$\Omega_B=0.06h$ where all the symbols have their usual
meaning.\newline

In section \ref{sec:galics} we describe the {\sc G}al{\sc ICS} model and
introduce the free parameters that it uses to control the star
formation efficiency, redshift dependence of the SFR, supernova
feedback and the efficiency with which gas that was once ejected by
the galaxy is re-accreted. In section \ref{sec:cal} we show the data
with which we compare our predictions and discuss their
uncertainties, mainly due to calibration issues. In section \ref{sec:qual} we determine the effect of varying the
free parameters and IMF upon the predicted relation and compare our
predictions with observations. In section \ref{sec:all} we then
present the results of using the best-fitting parameters for the entire
population of model galaxies and interpret the results physically. In section \ref{sec:conc} we draw our conclusions.

\section{The {\sc G}al{\sc ICS} Model}
\label{sec:galics} {\sc G}al{\sc ICS} is a semi analytic hybrid model of
hierarchial galaxy formation that combines the output of large
$N$-body cosmological simulations to track the evolution of baryonic
matter in galaxies through their dark matter haloes. The evolution of
galaxies is tracked using halo merging trees that follow the
hierarchial evolution of small objects at early times that may or
may not develop into larger ones through merging processes or
accretion of matter \citep{galics1}. {\sc G}al{\sc ICS} assigns a morphology to
a galaxy instantly after a merger based on the ratio of the bulge to
disc \textit{B}-band luminosities. In outline, as hot gas cools and falls into the centre of its dark matter halo,
it settles in a rotationally supported disc. The galaxies remain pure discs if their disc is globally
stable and they do not undergo a merger with another galaxy.
In the case where a significant merger occurs, we employ a recipe to distribute the stars
and gas in the galaxy between three different components in the resulting,
post-merger galaxy, the disc, the bulge, and a star-burst \citep[see][]{galics1}.
In the case of a disc instability, we simply transfer the mass of the gas and stars
necessary to make the disc stable to the burst component, and compute the properties of
the bulge/burst in a similar fashion to that described in \citet{galics1}.
The star-burst scale is $r_{\textrm{burst}}=0.1 r_{\textrm{bulge}}$, so that the characteristic
timescale for the star formation is shorter than in the bulge, hence
leading to a faster consumption of the gas. The star formation
rates are even higher than those in the discs, but have an \emph{instantaneous}
duration. The
\emph{burst} stellar population becomes part of
the bulge stellar population when the stars have reached an age of
100$\,$Myr.
Since we will only be using the model
predictions for high redshift it does not make sense to classify the
galaxies by local ($z=0$) standards and so here we consider all of
the predicted galaxies no matter the {\sc G}al{\sc ICS} assigned morphology.
\newline\newline The simulation models the universe as a box of comoving length of $150$ \textrm{Mpc}. Like any numerical
simulation, {\sc G}al{\sc ICS} has a finite baryonic mass resolution. The
minimum baryonic mass that we consider resolved in this simulation is $2\textrm{x}10^9M_\odot$, which is a factor of ten lower than the
fiducial {\sc G}al{\sc ICS} value (Hatton et al. 2003). The baryonic gas in galaxies is
initially primordial comprising of Hydrogen and Helium. The metal
content increases as time passes due to the synthesis of these
elements in stars during their lifetime and their subsequent release
into the inter-stellar medium (ISM) upon the stars death. A detailed
description of the entire {\sc G}al{\sc ICS} model may be found in
\citet{galics1} and an updated version (as far as the implementation
of the chemical evolution of O and Fe is concerned) of the model
that we use in this paper can be found in \citet{galics2}.\newline

The main novelty of the present version of \textsc{G}al\textsc{ICS} \citep{galics2} is the
implementation of a self-consistent treatment of the chemical
evolution with finite stellar lifetimes and both type Ia (SNIa) and type II (SNII)
supernovae ejecta. In practice, we follow the chemical evolution of
only four elements, namely H, He, O and Fe. This set of elements is
good enough to characterize our simulated galaxies from the
chemical evolution point of view as well as small enough in order to
minimize computational resources. Also, the reader should remember that
O is the major contributor to the total
metallicity.\newline\newline
We adopt the yields
from \citet{iwamoto} and references therein for both SNIa and
SNII.
{ The reader should note that a change in the stellar yields will introduce
a systematic offset of a few tenths of a dex in the model predictions \citep[see][]{tom,PM04}, hence it might leave room for some fine-tuning
for a suitable choice of stellar nucleosynthesis.
However, being only an offset, this change cannot create nor
modify the slope of the predicted relations.
%DTstart
But, most importantly, the successful calibration of our model with element ratios observed in Milky Way stars \citep[see][]{galics2}
does not allow significant modifications of the underlying stellar yields.}\newline\newline
The SNIa rate for a simple stellar population (SSP) formed at a given { time} is
calculated assuming the single degenerate scenario and the \citet{matrec} Delay Time Distribution (DTD).  The convolution of
this DTD with the SFR \citep[see][]{greg} gives the total SNIa rate.\newline\newline
Stars - and baryonic processes at the galactic scale that need finer detail -  are evolved between time-steps using sub-stepping of at least
1\,Myr. During each sub-step, stars release mass and energy into the
interstellar medium. In \textsc{G}al\textsc{ICS}, the enriched material released in the
late stages of stellar evolution is mixed to the cold phase, while the
energy released from supernovae is used to re-heat the cold gas and
return it to the hot phase in the halo. The rate of
mass loss in the supernova-driven wind that flows out of the disc is
directly proportional to the supernova rate (see below).\newline\newline
Throughout this work, we assume chemical homogeneity (instantaneous mixing),
such that outflows caused by feedback processes are assumed to have the same
metallicity as the inter-stellar medium, though in {reality the situation
cannot be captured by our simple recipe \citep{strick} and newly produced metals are more
likely to be ejected than the gas \citep[see, for example,][]{recc} }.
{Note that, due to the fine sub-stepping used for the stellar evolution, ejecta from SNII and the contributions of single low-mass stars is implemented without the need
to assume the instantaneous recycling approximation.}\newline

Below we focus on the main free parameters of the model that we will
change in order to study the build-up of the mass-metallicity
relation. A complete, detailed description of \textsc{G}al\textsc{ICS} and a compendium of its entire features may be found in \citet{galics1}.

\subsection{Initial Mass Function}

We use the Salpeter \citep{salp} IMF
\begin{equation}\label{eq:salp}\phi\!\left(m\right)\propto m^{-2.35}\end{equation} in our
simulations. The effect of changing the IMF is then investigated by
replacing it with the Kennicutt \citep{Ken} IMF,
\begin{equation}\label{eq:ken}\phi\!\left(m\right)\propto\left\{
\begin{array}{l l}
m^{-1.4}\,, & \quad \mbox{$ m < 1 M_{\odot} $}\\
m^{-2.5}\,, & \quad \mbox{otherwise}\\
\end{array} \right.\,, \end{equation} in section \ref{sec:qual}. We
take as the mass range $0.1 \le m \le 40 M_\odot$ for each IMF which
determines the normalisation. This is lower than adopted ($65M_\odot$) by \citet{amaze}; whose value would result in a higher O abundance on average of less than $0.1$ dex. We note here that an even higher cutoff (greater than $80M_\odot$) would lead to a change in the predicted O abundance of less than 0.3 dex, although
we must rely on the extrapolation of the nucleosynthetic yields available in the literature,
whereas the mass range $m<40M_\odot$ is where the O yields are more robust.
We have performed tests that show, as expected, that changes in the yields effect only the normalization of the Mass-metallicity relation and not
the slope. We therefore prefer to keep the same configuration as \citet{galics2}, where the chemical evolution
scheme (with a $40M_\odot$ upper limit) has been calibrated on observations of the Milky Way. Moreover, there are indications \citep{cesc}
that O production decreases with metallicity due to metal-dependent mass loss.
In light of the above caveats, the reader should keep in mind that some fine-tuning
of the normalization in the predicted mass-metallicity relation could be performed by acting on either the IMF or the O yields.

\subsection{Star Formation}

The star formation rate is given by

\begin{equation}\label{eq:SFR}\noindent
\textrm{SFR}=\alpha\frac{M_{\textrm{cold}}}{t_{\textrm{dyn}}}\left(1+z\right)^\beta\,,\end{equation}
where $M_{\textrm{cold}}$ is the mass of cold gas. The dynamical timescale
$t_{\textrm{dyn}}$ is defined as the time taken for matter at the half-mass
radius to reach either the opposite side of the galaxy (disc) or the
centre of the galaxy (bulge) and is given in \citet{galics1}. The
free parameter $\alpha$ sets the efficiency of star formation and
$\beta$ controls its redshift (time) dependence (if any). The\textsc{G}al\textsc{ICS} fiducial value of $\alpha$ \citep{galics1} is $0.02$ after \citet{guider} and others (see \citet{somer} tables 4 and 5) have used values in the range $0.01\le\alpha\le0.25$ with \textsc{G}al\textsc{ICS} restricted to the range $0.01\le\alpha\le0.1$ \citep{galics1}. In this work we vary $\alpha$ over the range $0.01\le\alpha\le0.05$. There is
observational evidence that the SFR was higher in the past
\citep{lilly,sper,helm,jun,Feulner} and so we focus on positive
values of $\beta$. Currently there is no ubiquitously accepted theoretical explanation for this. One possibility is that the
cosmic expansion means that galaxies were closer together in the past and
could thus interact more easily than at later times giving a higher merger rate than at present. There is, however, evidence that the number of star-forming galaxies at $z\approx 2$ is much higher than the predicted number of mergers \citep{conroy} indicating that much of the star formation is not merger driven. Some \citep{pipinosfr} argue that positive feedback from the central super-massive black hole during its early growth period can account for these observations and others that larger galaxies (that would have formed initially) have an overall higher cross section per unit mass \citep{ferreras} and thus accrete gas faster than smaller galaxies formed more recently. Successfully simulating a predicted mass-metallicity relation that closely matches the observed one may elucidate the mechanism by which galaxies have a higher SFR in the past. The \textsc{G}al\textsc{ICS} fiducial value of $\beta$ is $0$, however, using $\beta=0.6$ in order to incorporate
the effect of rapid accretion by cold flows, \citet{cat2} have managed to improve the fitting of the \textsc{G}al\textsc{ICS} predicted luminosity
function of Lyman-break galaxies to observations. Hence, we vary $\beta$ around this value. In section \ref{sec:discussion} we will discuss our predicted star formation rates in comparison with those that are observed in the galaxy samples with which we compare our predicted mass-metallicity relations.

\subsection{Supernova Feedback}
\label{sec:eps}

Massive stars will become type-II supernovae
that release energy into the ISM causing a fraction of the gas to be
ejected. If the energy is sufficient this ejected gas may leave the
galaxy inhibiting star formation. This process is known as supernova
feedback. We model the outflow rate using the formula from
\citet{silk03}

\begin{equation}\label{silk}\dot{m}=2\varepsilon\frac{E_{\textrm{SN}}\eta_{\textrm{SN}}}{v_{\textrm{esc}}^2}\textrm{SFR}\,.\end{equation}
where SFR is given in equation \ref{eq:SFR}. Here $\eta_{\textrm{SN}}$ is the
number supernovae per unit star forming mass, which we take as $7.4\textrm{x}10^{-3}$/$M_\odot$ and $E_{\textrm{SN}}$ is the
energy released by a single supernova assumed to have a constant
value of $10^{44}$J. The escape velocity $v_{\textrm{esc}}$ is given separately for bulges, discs and haloes in \textsc{G}al\textsc{ICS} I \citep{galics1}. The free parameter $\varepsilon$ determines the
efficiency at which gas from supernovae is injected into the ISM. The mass-loading factor is then equal to the factors pre-multiplying the SFR and hence $\varepsilon^{-1}$ can be though of as the efficiency of mass-loading. The \textsc{G}al\textsc{ICS} fiducial value is $0.3$ and others have used values in the range $0.05\le\varepsilon\le0.2$ (see \citet{somer} tables 4 and 5). \citet{cat2} have found that $\varepsilon=0.2$ is the best-fitting value if \textsc{G}al\textsc{ICS} is used to predict the luminosity
function of Lyman-break galaxies. Hence, we initially vary $\varepsilon$ in the range $0\le\varepsilon\le0.3$.

\subsection{Ejection of Matter from the Halo}

Type-II supernova and other galaxy forming processes may lead to the
ejection of gas and metals from the halo. These individual processes
are difficult to model \citep{galics1} and the {\sc G}al{\sc ICS} model
accommodates them by storing all ejected gas and metals ejected from
the halo in a `reservoir' from which they may or may not be
re-accreted at a later time. When matter is accreted from extra
galactic sources a certain fraction is primordial and the rest is
drawn from the reservoir. The efficiency of re-accretion from the
reservoir is characterized by the parameter $\zeta$, the halo
recycling efficiency. If $\zeta=0$ then material ejected from the
halo is lost permanently. If $\zeta=1$ then all matter accreted is
drawn from the reservoir until it is depleted at which point any
further gas accreted is assumed primordial. Recently, \citet{opp} have studied the effects of recycling on the SFRs and stellar mass function of galaxies in cosmological hydrodynamical simulations and have found that it is the dominant factor in galaxy growth at $z\le1$, ejecta being the important factor at $z\ge2$. Hence, in this work we focus on the supernova feedback efficiency and do not vary $\zeta$, holding it constant at the \textsc{G}al\textsc{ICS} fiducial value, $0.3$, so that during any gas re-accretion process, 30\% is drawn from the reservoir (provided it is not depleted), the remainder being of primordial metallicity.

\section{Data}
\label{sec:cal}

The gas metallicity must be determined using strong line metallicity
diagnostics \citep{amaze} which relies on the fact that the ratio of
various strong optical emission lines depend upon the gas
metallicity in a known manner. Thus these ratios must be calibrated
against metallicity. Calibrations have only been performed in narrow
metallicity ranges. These calibrations are often inconsistent with
each other and can lead to different metallicity offsets of up to
$0.7$ dex and the difference in the shape of the curve is often
large \citep{kew08}. Since data measured at different redshifts may
be measured using different optical lines it is essential to ensure
a correct intercalibration between the data fits so that the correct
evolution of the relation can be seen. These intercalibration issues are tackled by \citet{Nagao} and by
\citet{amaze},
who take a large sample of local galaxies spanning a wide range of metallicities
($7.2< 12+\log(O/H)<9.2$, accurately measured by using both the electron temperature
method
and photoinoization models) and cross-calibrate the various strong line ratio
diagnostics
on the same metallicity scale.\newline

We take as our observed trend the AMAZE \citep{amaze}
mass-metallicity relation \begin{equation}\label{eq:cal}12+\log\left(O/H\right)=-0.0864\left(\log
M_\star-\log M_\mathrm{0}\right)^2+K_\mathrm{0}\,,\end{equation}where $\log M_\mathrm{0}$ and
$K_\mathrm{0}$ are free parameters that must be determined at each redshift
to obtain the best-fitting to the observed data and are shown in Table
\ref{tab:cal}. The calibration constants at $z=0.07$ were derived
using the data from \citet{kew08}, the constants at $z=2.27$ were
found using the data from
\citet{erb06} and the constants corresponding to $z=3.54$ were calculated using data from \citet{amaze}.

\begin{table}\centering
\begin{tabular}{ l l l}
   $z$& $\log M_\star$ & $K_0$ \\ \hline
   $0.07$& $11.18$ &$9.04$\\
   $2.27$& $12.38$ &$8.99$\\
   $3.54$ &$12.87$ & $8.90$\\ \hline
 \end{tabular}\caption{Calibration constants used in equation \ref{eq:cal} as a function of redshift
from \citet{amaze}.}\label{tab:cal}
\end{table}

\section{Results}
\label{sec:qual}

In this section we test the effects of changing the parameters, in
order to identify those that may improve the agreement with
observations. A more quantitative discussion will involve only these
parameters. Initially all qualitative results were obtained using
the Salpeter IMF (equation \ref{eq:salp}) however the effect of changing
to the Kenicutt IMF (equation \ref{eq:ken}) is investigated in
section \ref{sec:kenimf}.

\subsection{\label{sec:vary} Variation of Parameters}

\subsubsection{Star Formation Efficiency} \label{sec:alpha}

\begin{figure}
  \begin{center}
      \subfigure[$\alpha=0.01$]{\includegraphics[width=8 cm,height=8 cm]{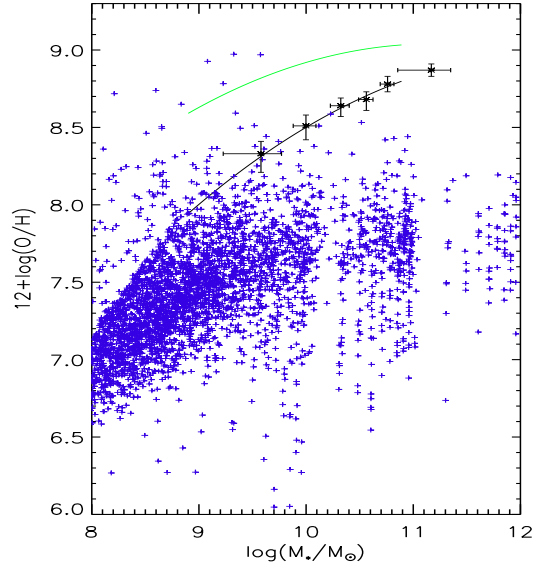}}
      \subfigure[$\alpha=0.05$]{\includegraphics[width=8 cm,height=8 cm]{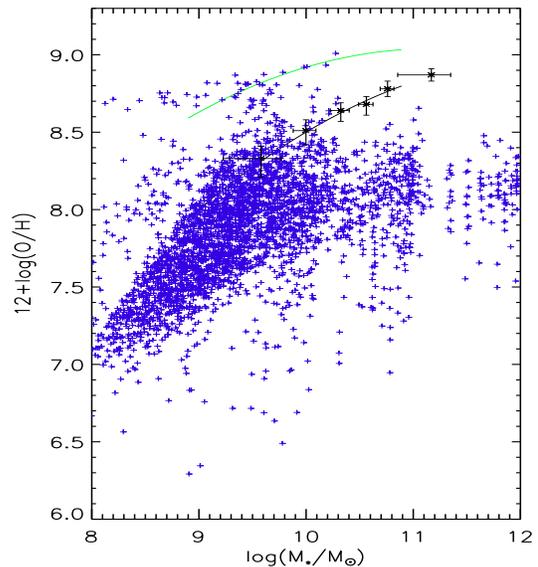}}\caption{Response of the predicted mass-metallicity relation to a change in the star
      formation efficiency for galaxies at redshift $2.27$. The blue crosses represent an individual galaxy with stellar mass and metallicity predicted by the model and the black asterisks show the data from
      \citet{erb06}. The green (upper) line shows the AMAZE best-fitting line for $z=0.07$ and the black (lower) line is the same curve fitted to the data at $z=2.27$.}
\label{fig:alpha1}
  \end{center}
\end{figure}

\begin{figure}
  \begin{center}
      \subfigure[$\alpha=0.01$]{\includegraphics[width=8 cm,height=8 cm]{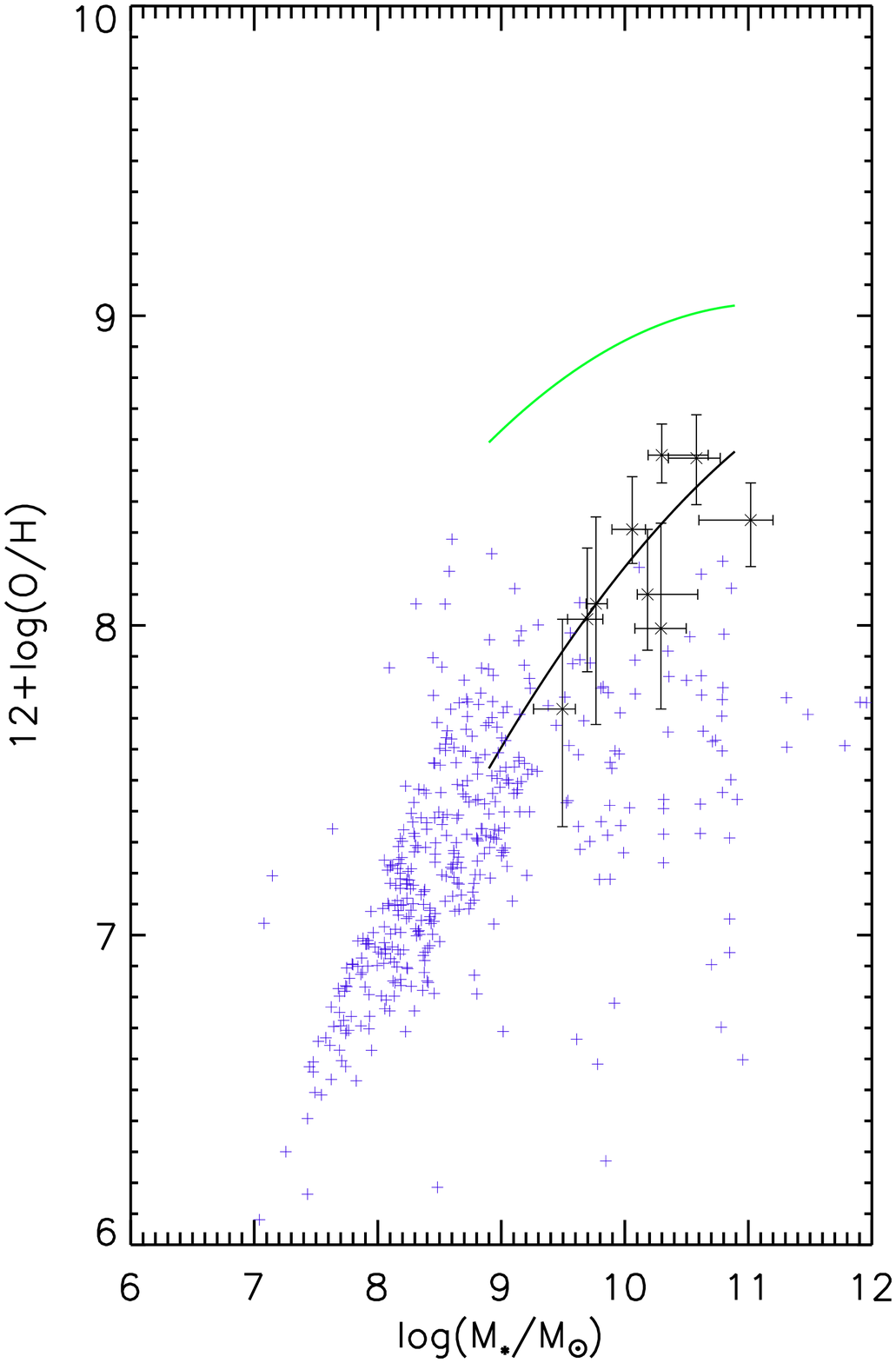}}
      \subfigure[$\alpha=0.05$]{\includegraphics[width=8 cm,height=8 cm]{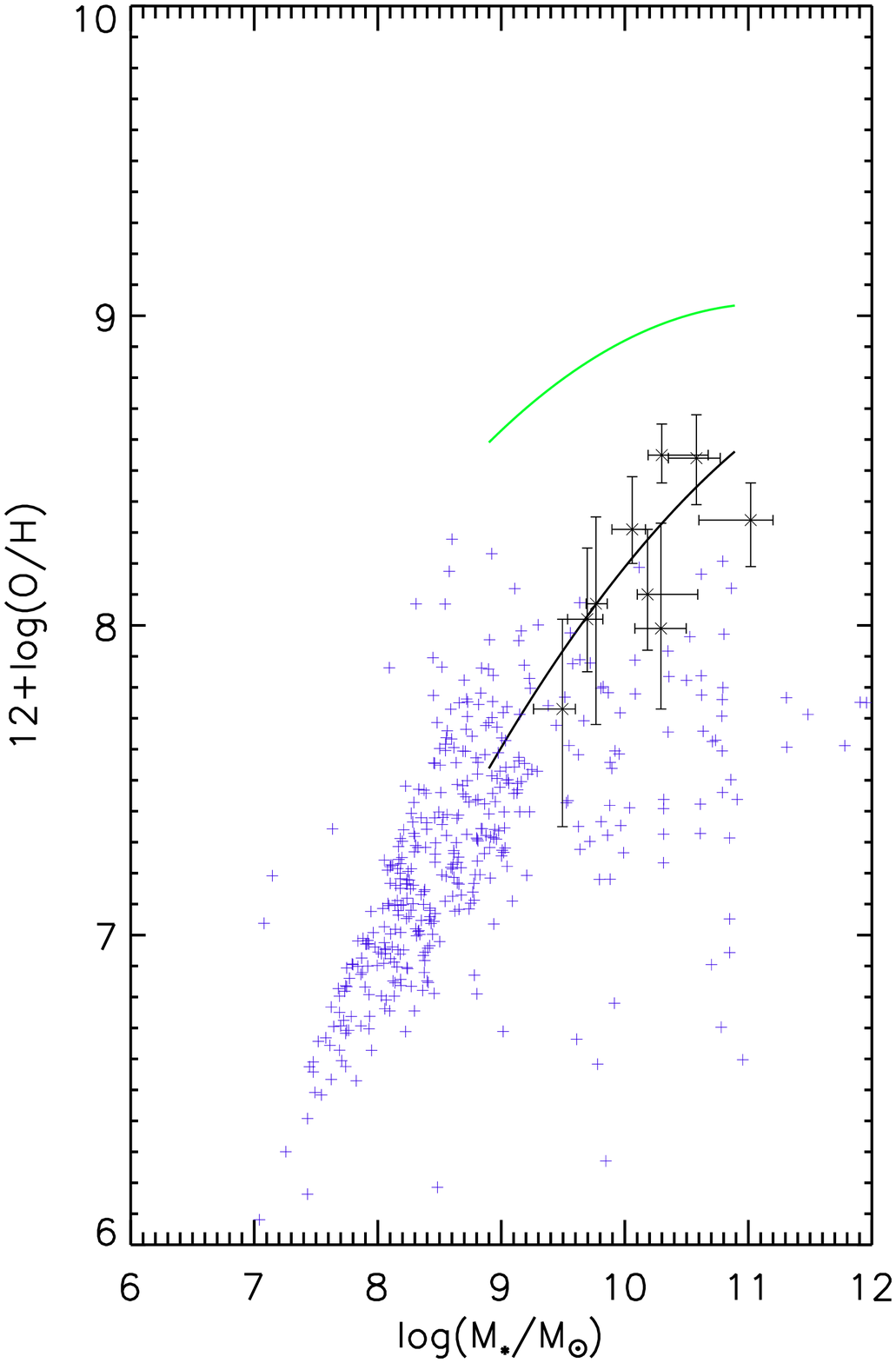}}\caption{Response of the predicted mass-metallicity relation to a change in the star
      formation efficiency for galaxies at redshift $3.54$. The blue crosses represent an individual galaxy with stellar mass and metallicity predicted by the model
      and the black asterisks show the data from \citet{amaze}. The green (upper) line shows the AMAZE best-fitting line for $z=0.07$ and the
black (lower) line is the same curve fitted to the data at
$z=3.54$.}\label{fig:alpha2}
  \end{center}
\end{figure}

Holding $\beta$, $\epsilon$ at their fiducial values
\citep{galics1}, the parameter $\alpha$ was varied over the range
$0.01<\alpha<0.05$ to investigate the effects of changing the star
formation efficiency on the predicted mass metallicity relation.
Fig. \ref{fig:alpha1} shows the output for galaxies at redshift
$2.27$ plotted as $12+\log\left(O/H\right)$ Vs. $\log\!\left(
M_\star/M_\odot\right)$ where $M_\star$ is the stellar mass. Also plotted are the observations at $z=2.27$ from \citet{erb06} as
well as the calibrated best-fitting trend (equation \ref{eq:cal}) at
redshift $2.27$ and $0.07$. The
corresponding plot for galaxies at redshift $3.54$ is shown in
Fig. \ref{fig:alpha2}.
We note that, although the minimum baryonic mass is $2\textrm{x}10^{9}M_\odot$, it
 is possible to have galaxies whose stellar mass is less than this in the sample provided that their baryonic mass exceeds this minimum value.
The plots show that increasing $\alpha$ has the effect of spawning a
similar number of galaxies (at the same redshift) that have on
average a higher mass and metallicity. Increasing $\alpha$ increases
the SFR in direct proportion thus in the same Hubble time we have
more stars formed and a larger proportion at a later stage in their
life and so the stellar mass and metallicity is increased. We note
from Fig. \ref{fig:alpha2} that the observed mass-metallicity
relation \citep{amaze} would be best fitted using a low star formation
efficiency for low-mass galaxies and a high star formation
efficiency for high-mass galaxies. This supports the findings of
\citet{amaze} who argue that low-mass galaxies are characterized by
low star formation efficiencies inhibiting chemical evolution. We will return to this issue later in section \ref{sec:discussion}.
\newline\newline
From a formal point of view, if we quantify
the agreement between model predictions and data in terms of the $\chi^2$, we have values that monotonically
decrease with increasing $\alpha$, simply because the normalization of the predicted
mass-metallicity relation increases and, on average, more model galaxies are in better
agreement with the data. This trend, however, has the effect of predicting too many
small galaxies at $z\sim3$ that exhibit the metallicity of a typical $z\sim2$ galaxy of the
same mass. This trend is already present at $\alpha=0.06$
without inducing any improvement of the slope of the predicted relation.
When discussing the yields, we showed that we adopt quite a conservative value for the O production,
therefore we believe that values for  $\alpha>0.05$ should be avoided
even if they lead to a low value $\chi^2$. Also, the reader should note here that during the preparation of the manuscript, several
papers have introduced a more robust way to explore the parameter space and statistically handle the comparison
between model predictions and observations \citep[see, for example,][]{bower,lu}.
In particular, \citet{lu} argue that the procedure adopted here (namely varying only
one parameter at a time) does not allow one to uniformly explore the parameter space and that
the ``best-fitting by eye'' values do not always coincide with the point that has the maximum likelihood
in the parameters multi-dimensional phase space. On the other hand, the procedure \citet{lu} advocate
may lead to a formal best-fit parameter set that is either unphysical or difficult to explain
from the theoretical point of view. Therefore some priors on the parameters have to be adopted.
The present work aims at probing the sensible range for some of those.

\subsubsection{Redshift Dependence of the
SFR}\label{sec:beta}

\begin{figure}
  \begin{center}
      \subfigure[$\beta=0$]{\includegraphics[width=8 cm,height=8 cm]{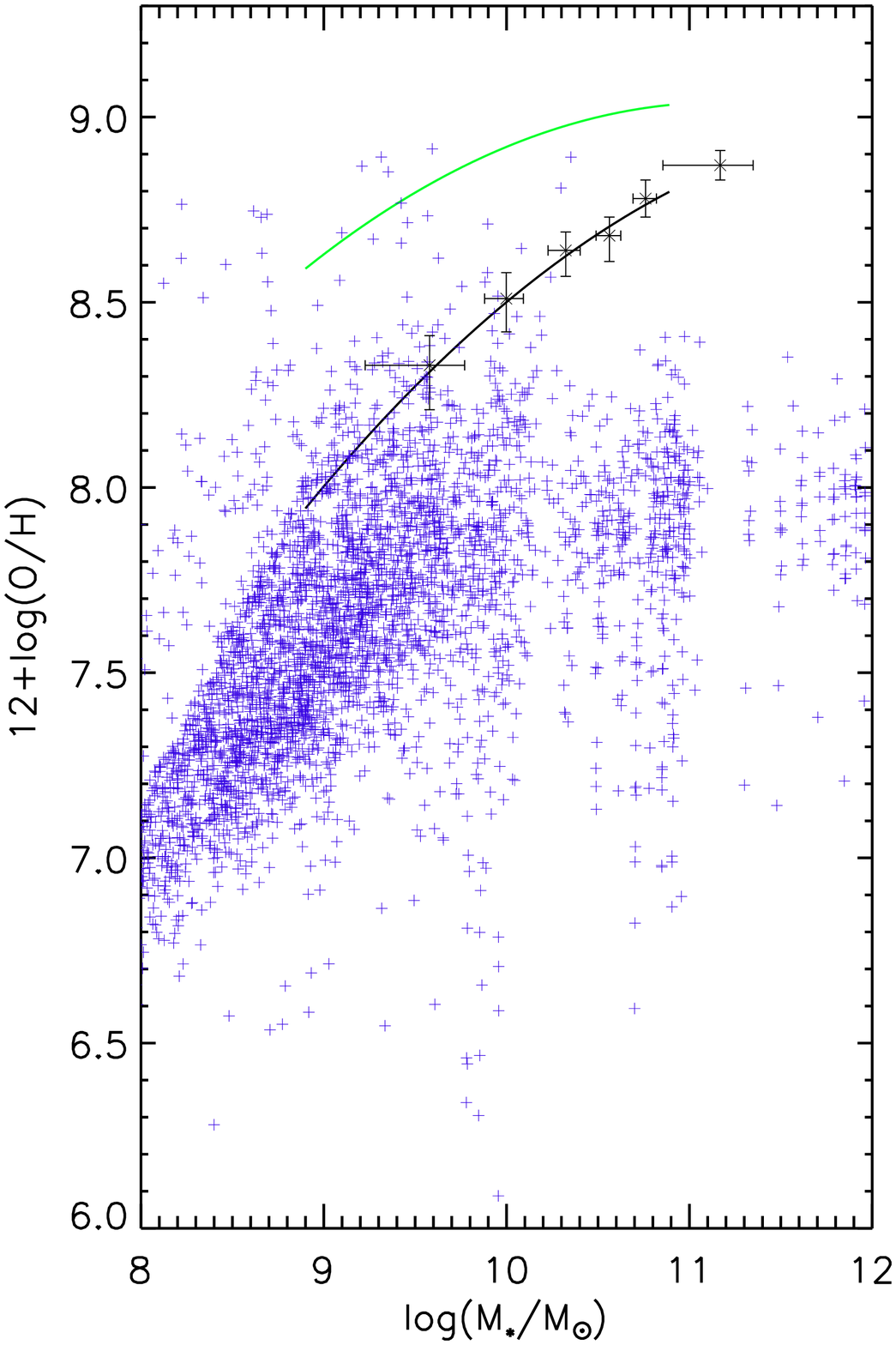}}
      \subfigure[$\beta=1.25$]{\includegraphics[width=8 cm,height=8 cm]{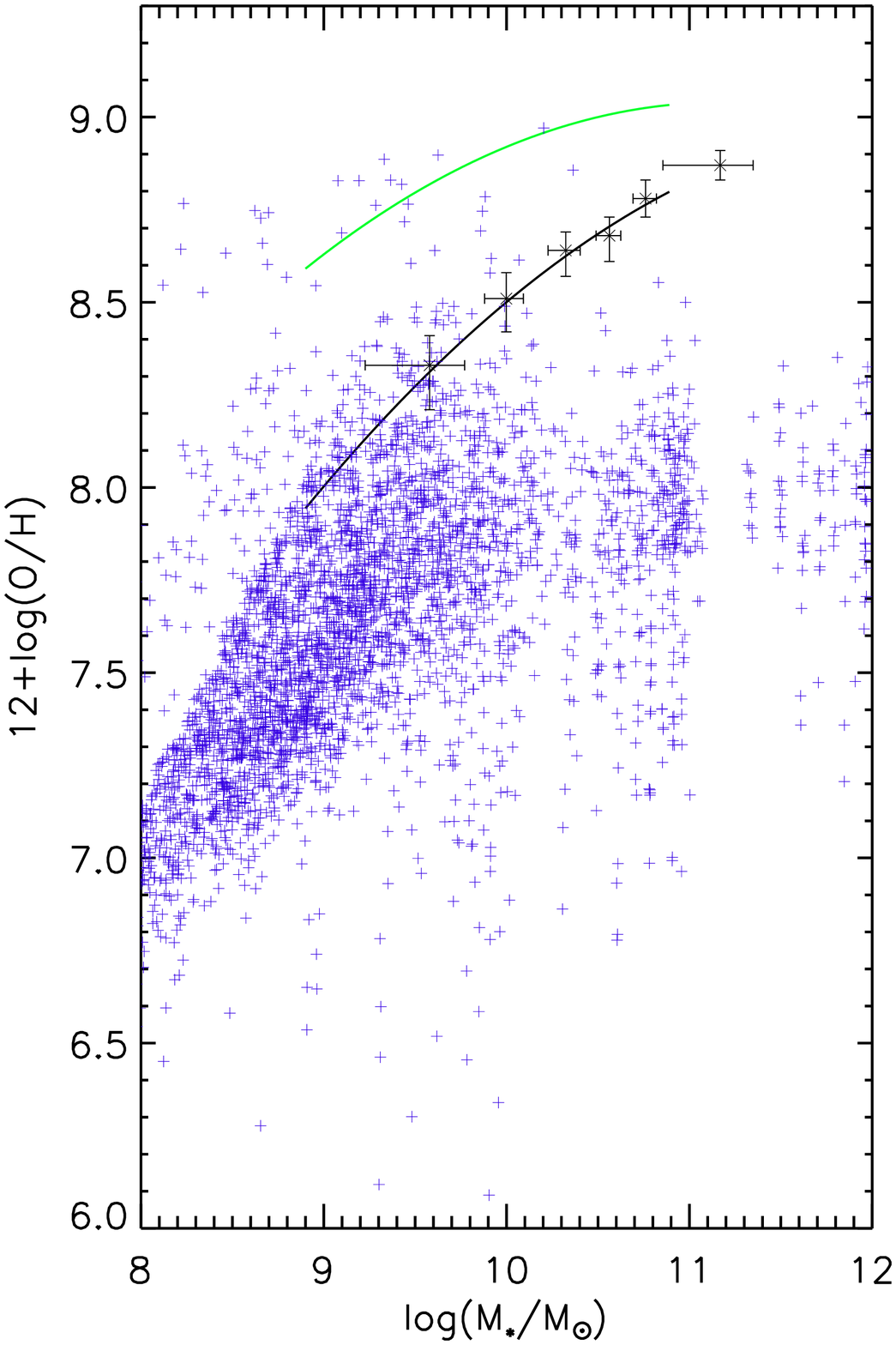}}
      \caption{Response of the predicted mass-metallicity relation to a change in the redshift dependence of the star
      formation rate for galaxies at redshift $2.27$. The blue crosses represent an individual galaxy with stellar mass and metallicity predicted by the model and the black asterisks show the data from \citet{erb06}.
      The green (upper) line shows the AMAZE best-fitting line for $z=0.07$ and the black (lower) line is the same curve fitted to the data at $z=2.27$.}\label{fig:beta1}
  \end{center}
\end{figure}

\begin{figure}
  \begin{center}
      \subfigure[$\beta=0$]{\includegraphics[width=8 cm,height=8 cm]{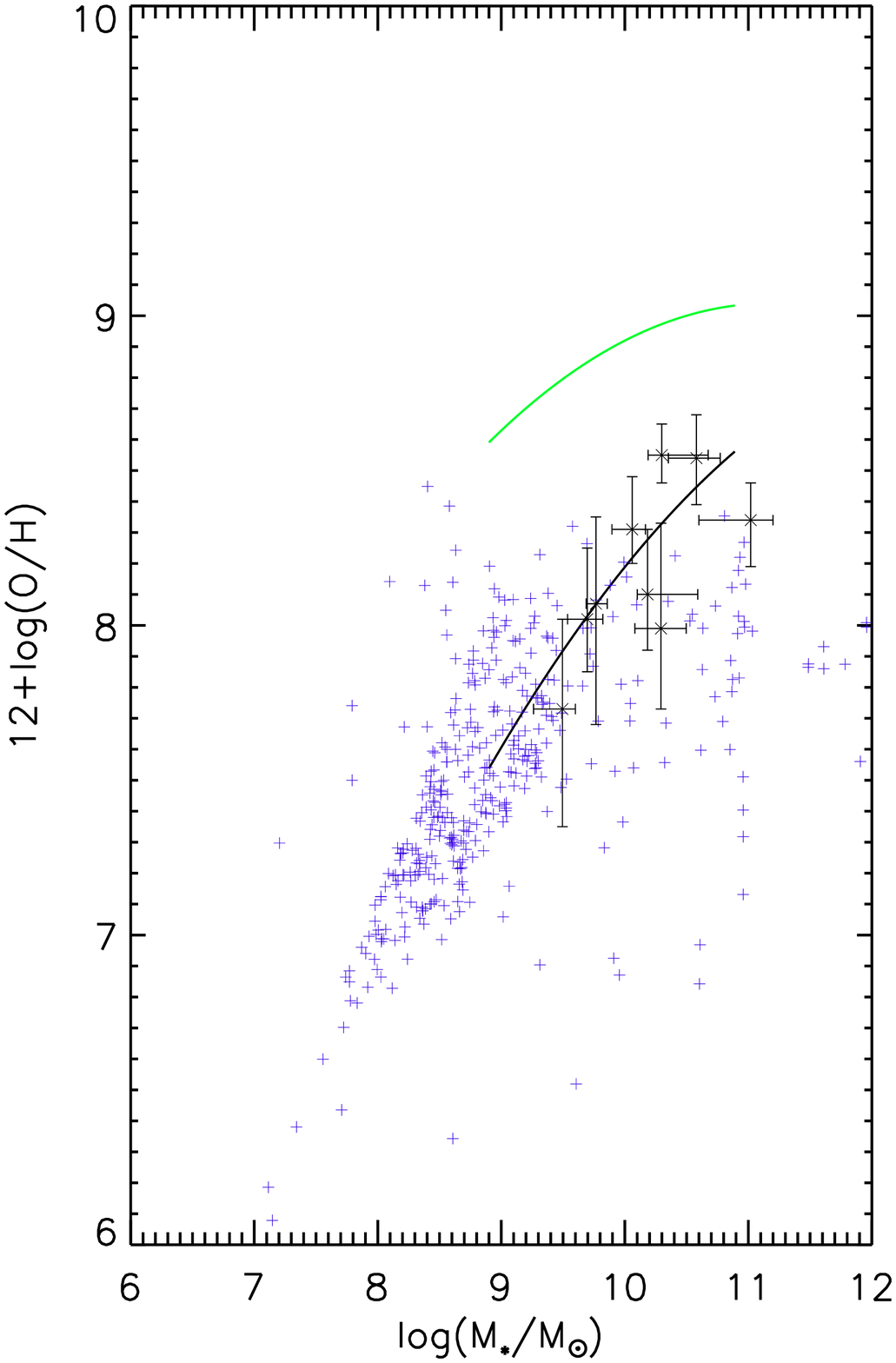}}
      \subfigure[$\beta=1.25$]{\includegraphics[width=8 cm,height=8 cm]{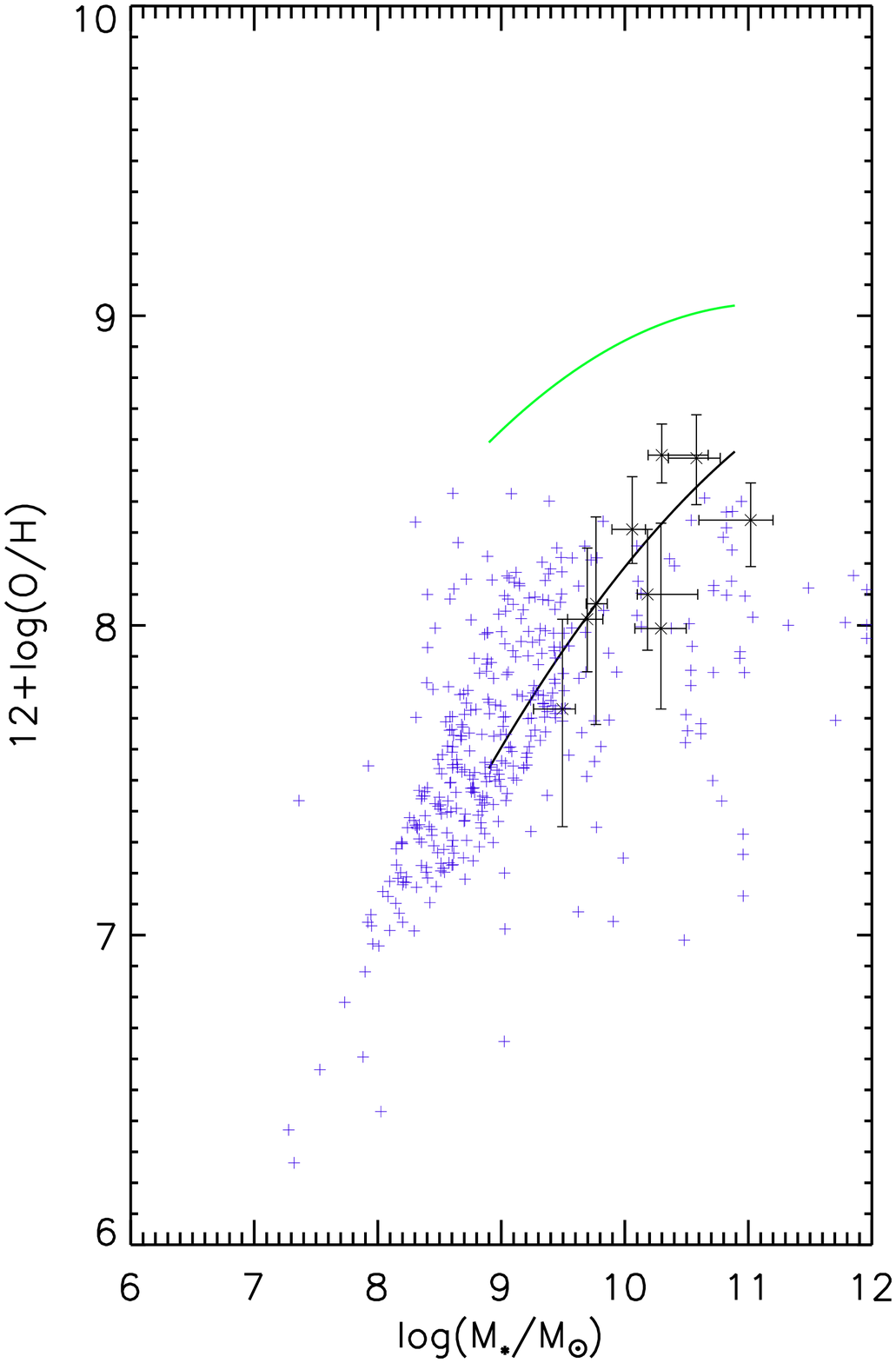}}\caption{Response of the
      predicted mass-metallicity relation to a change in the redshift dependence of the star
      formation rate for galaxies at redshift $3.54$. The blue crosses represent an individual galaxy with stellar mass and metallicity predicted by the model and the black asterisks show the data from \citet{amaze}.
      The green (upper) line shows the AMAZE best-fitting line for $z=0.07$ and the black (lower) line is the same curve fitted to the data at $z=3.54$.}\label{fig:beta2}
  \end{center}
\end{figure}

With $\alpha$, $\epsilon$ and $\zeta$ held constant at their
fiducial values $\beta$ was varied over the range $0<\beta<1.25$.
The {\sc G}al{\sc ICS} predictions for galaxies are shown in Fig.
\ref{fig:beta1} ($z\approx2.27$) and Fig. \ref{fig:beta2}
($z\approx3.54$). Inspection of the plots show that $\beta$ has a distinguishable effect
on the distribution of the mass and metallicities of the galaxies. At any given
redshift increasing $\beta$ preserves (approximately) the number of
galaxies predicted but the distribution favours more galaxies with a
higher stellar mass and metallicity whilst preserving the trend
which is a similar shape to the AMAZE calibration curve (equation
\ref{eq:cal}). This follows from the fact that at constant redshift
increasing $\beta$ results in the term $\left(1+z\right)^\beta$ in
equation \ref{eq:SFR} being larger and thus it acts in a similar
manner to $\alpha$.  The same number of mergers are predicted so
that the mass distribution is similar but not identical since the
SFR does play a role in determining the mass of individual galaxies \citep{wei07}.\newline\newline
As in the previous section a simple $\chi^2$ analysis would return
lower results for increasing $\beta$, whereas we believe that the values
for $\beta$ should not be higher than 1.

\subsubsection{Supernova Feedback Efficiency
\label{sec:varyeps}}

\begin{figure}
  \begin{center}
      \subfigure[$z=2.27$\label{fig:epsz2}]{\includegraphics[width=8 cm,height=8 cm]{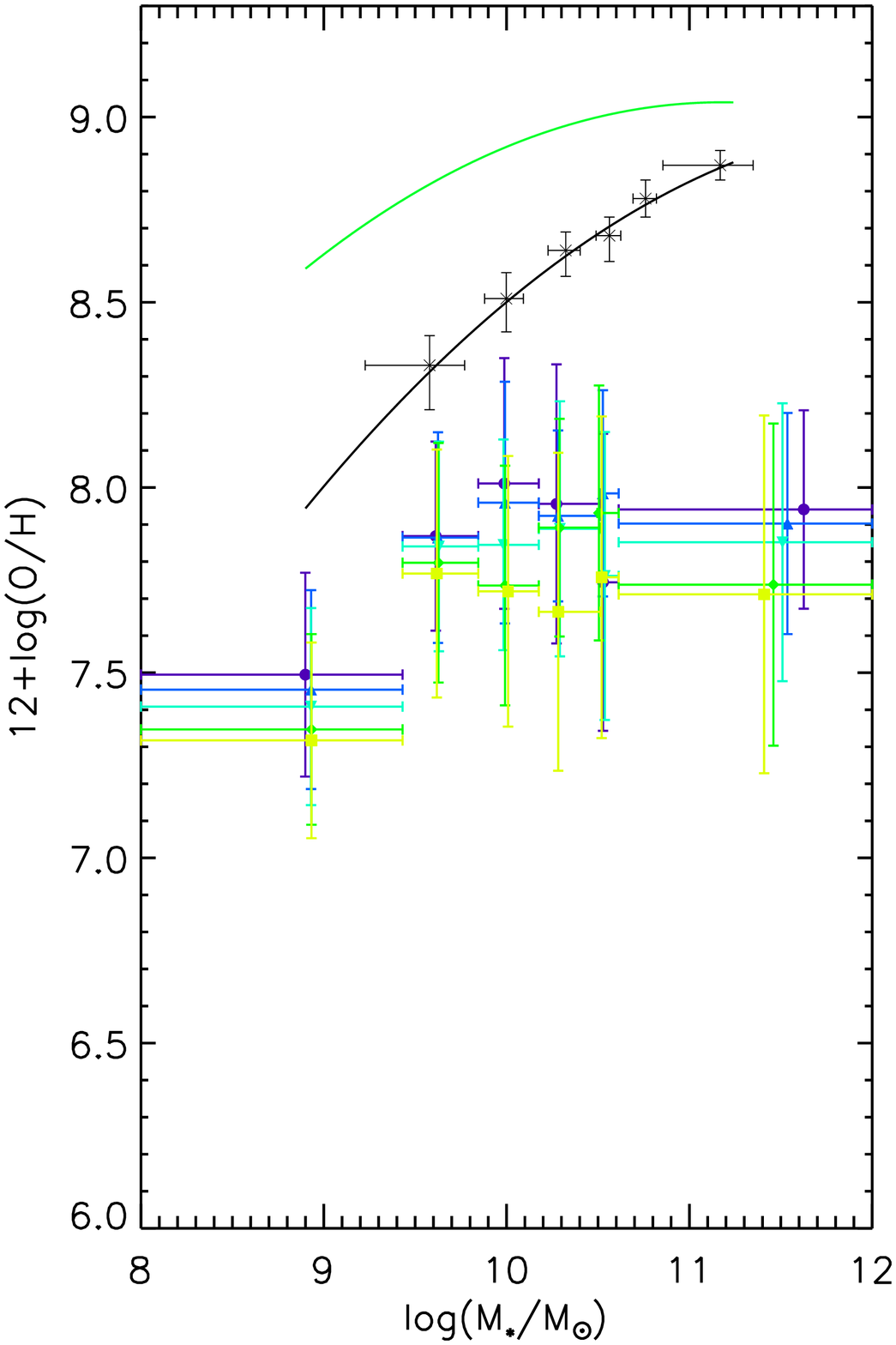}}
      \subfigure[$z=3.54$\label{fig:epsz3}]{\includegraphics[width=8 cm,height=8 cm]{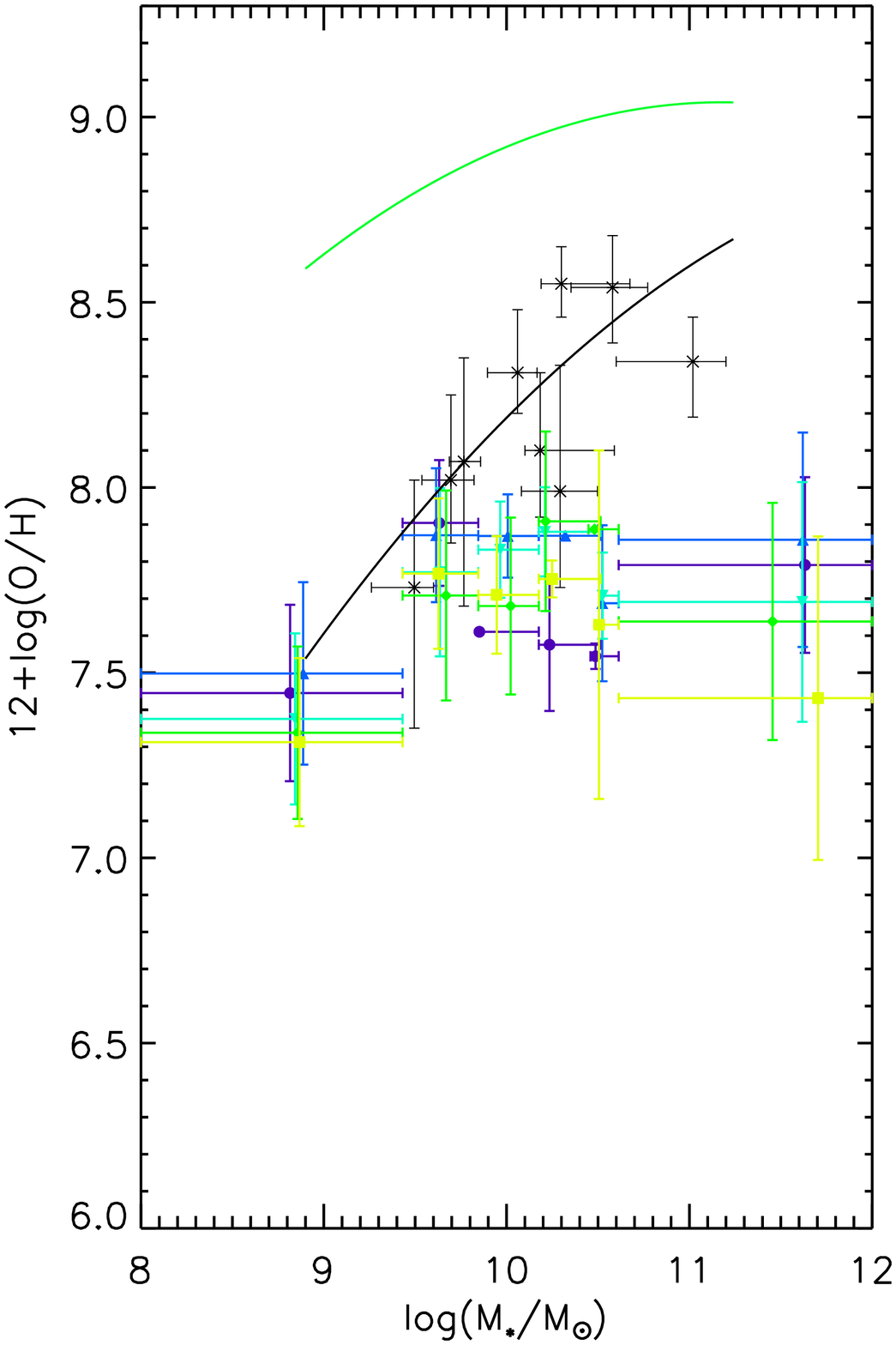}}
      \caption{The change in the average mass-metallicity relation when $\varepsilon$ is varied at redshift $2.27$ and $3.54$. The purple circles, dark blue upward-facing triangles, light blue downward-facing triangles, green diamonds and yellow squares show the average mass and metallicity in each of the six mass bins used by \citet{erb06} when $\varepsilon=0.1\,,0.3\,,0.5\,,0.8$ and $1.0$ respectively. The horizontal bars show the range of the mass bins and the vertical bars show one standard deviation in the metallicity. The black asterisks and line show the data from \citet{erb06} (Fig. \ref{fig:epsz2}) or \citet{amaze} (Fig. \ref{fig:epsz3}) and the black (lower) line shows the AMAZE best-fitting line for the observed relations at the respective redshifts. The green (upper) line shows the AMAZE best-fitting line for the observed relation at $z=0.07$.}\label{fig:epsvary}
  \end{center}
\end{figure}

With $\alpha$ and $\beta$ held constant at their fiducial values $\varepsilon$ was varied over the range $0.1\le\varepsilon\le1.0$. Following the discussion in section \ref{sec:eps}, this range is investigated in order to test the model however we do not use excessively large values in order to artificially fit relations hereafter. The plots in Figs. \ref{fig:epsz2} and \ref{fig:epsz3} show the mean mass and metallicity predicted in each of the six mass bins used by \citet{erb06} at $z=2.27$ and $z=3.54$ respectively. Fig. \ref{fig:epsz2} shows that the effect of increasing epsilon at $z=2.27$ is to lower the average metallicity in each bin whilst preserving the overall shape of the relation. At higher values of $\varepsilon$, more gas is ejected from the galaxy thereby reducing the average metallicity in each bin. Thus the effect of changing $\varepsilon$ is to alter the offset of the relation without altering the slope. This effect is comparable to the one predicted by \citet{findave08} with the difference that we have a self consistent chemical evolution that includes the finite lifetimes of both SNIa and SNII. At $z=3.54$, Fig. \ref{fig:epsz3} shows that changing the value of $\varepsilon$ has very little effect on the relation, especially in the low mass regime, and basically preserves the shape of the relation. \newline\newline
\citet{Mann}, whose sample of LBGs at $z\sim3$ (discussed in the introduction), have found that the effective yield (the amount of metals synthesised and retained within the ISM per unit stellar mass) decreases with increasing stellar mass for galaxies in their sample suggesting that galactic outflows cannot account for the shape of the mass-metallicity relation since their power is diminished in more massive galaxies and thus they cannot be responsible for the decreasing effective yields. Using chemical evolution models for galaxies of varying morphological types \citet{calura} have found that the relation arises naturally, regardless of the morphology, if the SSFR is larger in more massive galaxies and that galactic outflows are not needed to explain the relation. As noted in the introduction, none of the models that focus solely on different outflow processes only have been able to fit the observed relation at $z=3.54$ and, taken together, these plots imply that outflows are only important in determining the low redshift relation whereas at higher redshifts, another mechanism is responsible for generating this relation, which may explain this lack of predictive power at higher redshifts. Considering these recent results, the findings of this section imply that it is the SFR-mass relation that generates the observed mass-metallicity relation and determines the slope at high redshift with outflows being important only for the low redshift properties.

\subsection{Variation of the IMF}
\label{sec:kenimf}

\begin{figure}
  \begin{center}
      \subfigure[$z=2.27$\label{fig:changea}]{\includegraphics[width=8 cm,height=8 cm]{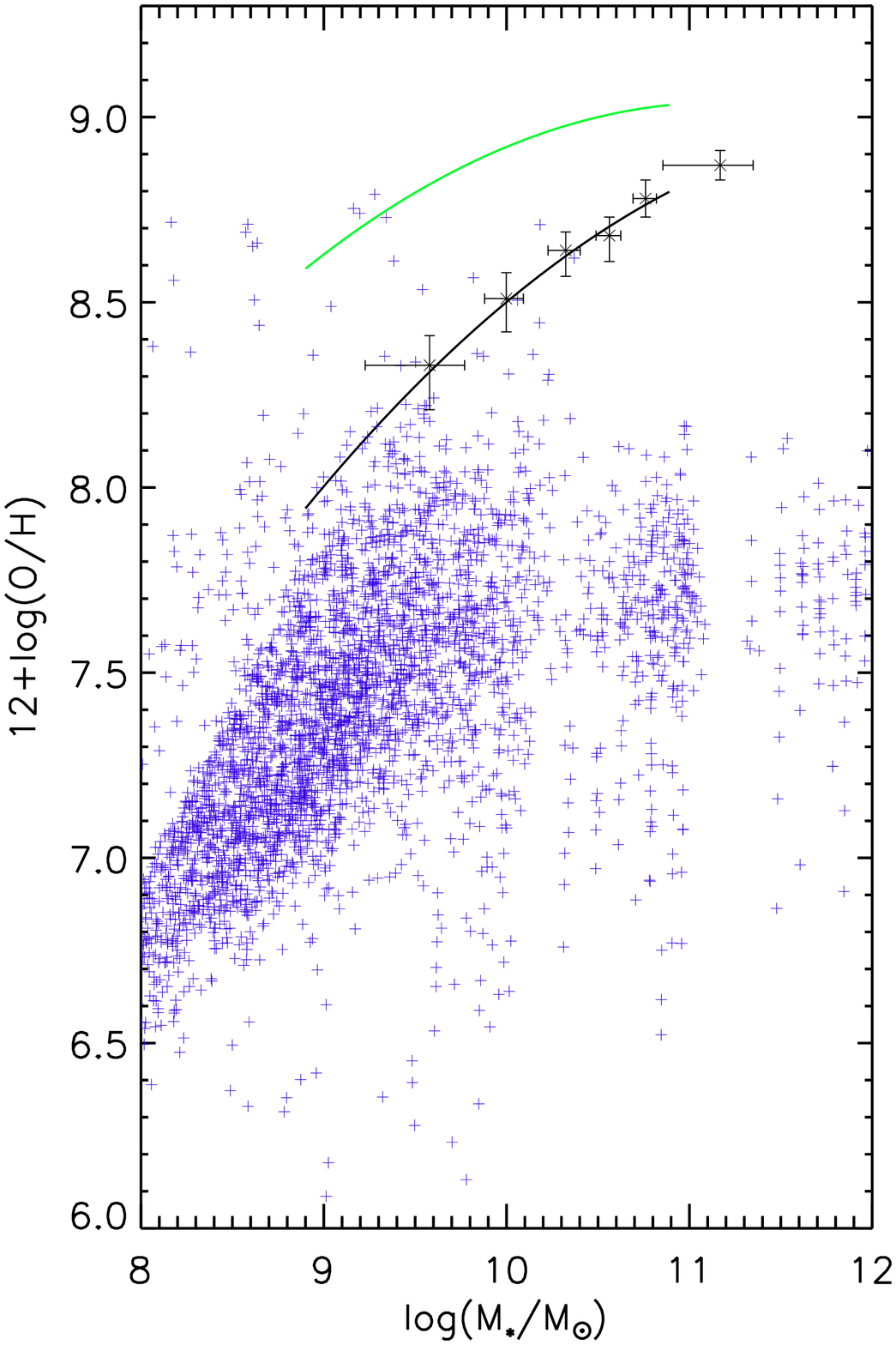}}
      \subfigure[$z=3.54$\label{fig:changeb}]{\includegraphics[width=8 cm,height=8 cm]{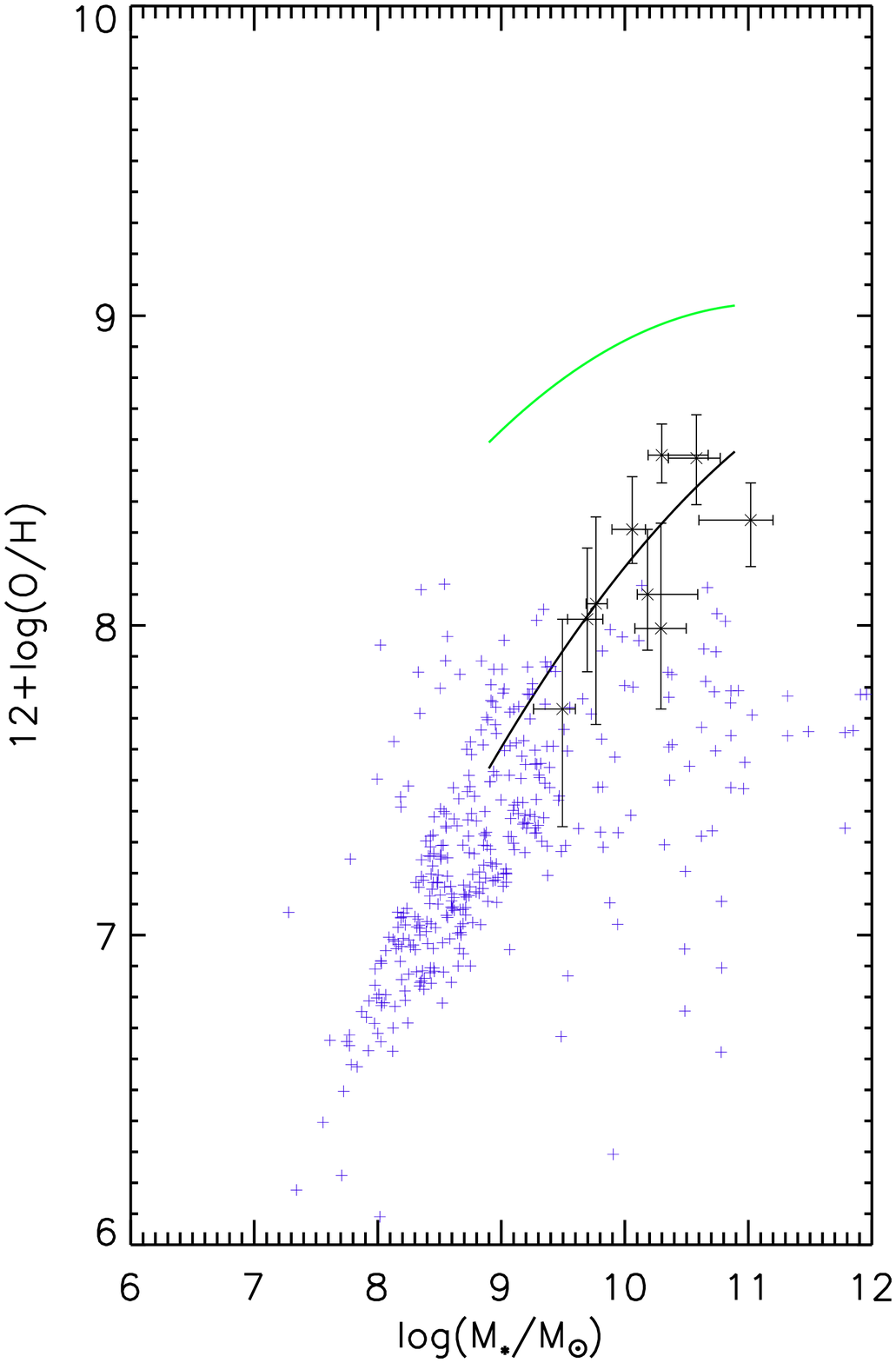}}
      \caption{Predicted mass-metallicity relation using the Kennicutt IMF with the free parameters held at their fiducial values.
      The blue crosses represent an individual galaxy with stellar mass and metallicity predicted by the
      model and the green (upper) line shows the AMAZE best-fitting line for $z=0.07$. The black asterisks show the data from
      \citet{erb06} (\ref{fig:changea}) or the data from \citet{amaze} (\ref{fig:changeb}) and
      the black (lower) line shows the AMAZE best-fitting line for $z=2.27$ (\ref{fig:changea}) or $z=3.54$ (\ref{fig:changeb}).}\label{fig:imfchange}
\end{center}\end{figure}

Using the free parameters held at their fiducial values the IMF was
changed to the Kennicutt IMF (equation \ref{eq:ken}). Fig.
\ref{fig:imfchange} shows the predicted relation for galaxies at
redshifts $2.27$ and $3.54$. Comparing Fig. \ref{fig:imfchange} to
Figs. \ref{fig:alpha1}-\ref{fig:beta2} it is clear that both IMFs
predict similar shaped relations at both redshifts. As a matter of fact, the
Kennicutt IMF predicts lower values than both Salpeter and
observation at $z=2.27$. At $z=3.54$ both IMFs predict similar
distributions that are both in the same region as observed. Only the
Salpeter IMF produces a distribution that has many galaxies with
similar masses and metallicities to observations at both redshifts,
which indicated that it may have been able to reproduce the observed
relation given the right values of the free parameters. For
this reason only the Salpeter IMF was investigated quantitatively to
see if it could reproduce the observed relation at both redshifts.

\section{Discussion}\label{sec:discussion}

Examination of the plots in Figs. \ref{fig:alpha1} to \ref{fig:beta2} reveals that, no
matter the value of any parameter, there is a definite trend in the
predicted distribution of stellar mass and metallicity. This is the
known trend of increasing metallicity with stellar mass that follows
a shape similar to the AMAZE calibration curve (equation
\ref{eq:cal}) at both redshifts. The plots also show a scatter that is comparable to the scatter seen
in the data from \citet{trem04} at $z\approx0$ independent
of the value of any of the free parameters. Finally most of the
galaxies have metallicities that lie below the data from
\citet{trem04} ($z\approx0$) and the AMAZE best-fitting curve for
$z=0.07$ indicating that most of the galaxies had a lower
metallicity in the past than today. This is consistent with stellar
and galactic chemical evolution theory \citep{pag} and is an
indication that the model simulates the physics correctly.\newline

\label{sec:fit} The results of section \ref{sec:vary} showed that
the model could reproduce the observed mass-metallicity relation
and suggested that a good fit to the current observations could be
achieved by a suitable choice of parameters and IMF. Section
\ref{sec:varyeps} revealed that the effects of
varying $\varepsilon$ were much smaller than varying
$\alpha$ and $\beta$ hence only these two parameters were varied.
Predictions were generated using parameters in the range
$0.01\le\alpha\le0.05$ and $0\le\beta\le1.25$ with $\varepsilon$ and
$\zeta$ held constant at their fiducial values.\newline

It was found that the parameters that best match the observed
relations were $\alpha=0.02$ and $\beta=1.0$.
\label{sec:all}
The plots in Fig. \ref{fig:best} show the predicted relations at
both redshifts using the best-fitting parameters found in section
\ref{sec:fit}. Fig. \ref{fig:asala} shows that the observed
relation is not predicted at $z=2.27$, instead the predicted
distribution has a large scatter and shows a similar shaped trend to
but lies mainly below the observations of \citet{erb06}. Fig.
\ref{fig:asalb} shows that the observed relation \citep{amaze} is
predicted at $z=3.54$ with a comparable scatter. At both redshifts
the majority of predicted galaxies lie below the observations of
\citet{trem04} and the AMAZE $z=0.07$ best-fitting
relation (equation \ref{eq:cal}).
At both redshifts we predict a flatter relation than observed. This effect is
particularly clear at $z=2.27$ where none of the predicted massive galaxies are consistent
with the observed ones even taking into account the errors on the latter. The situation
is less extreme in the high redshift case, nevertheless it may be a symptom of some missing physics
in the model. We will thoroughly discuss such problems in the rest of this section.
To gain better insights on the model galaxies it is worth, however, first having a look at the predicted SFRs,
since the predicted flattening in the mass-metallicity relation may come from issues with the manner in which \textsc{G}al\textsc{ICS} predicts
them.
Both plots show the predicted galaxies stratified by SFR into the three bins $\textrm{SFR}<10 M_\odot \textrm{yr}^{-1}$, $10\le\textrm{SFR}<30 M_\odot \textrm{yr}^{-1}$ and $\textrm{SFR}\ge30 M_\odot \textrm{yr}^{-1}$. To some extent, the predicted SFRs should only be compared with each other and any direct comparison with experimentally observed values should be treated equivocally. In fact, \textsc{G}al\textsc{ICS} predictions are average values over the entire galaxy
and over the timestep, whereas observations look at more central regions (see below) and yield an instantaneous SFR value. We can thus gain more insight from the comparison of the relative differences and evolution in the predicted and observed SFRs than their numerical values. Indeed, our predicted SFRs (see Fig. \ref{fig:srf} below) are quite lower than those observed by \citet{amaze} who find an average SFR of $\sim$100$M_\odot$yr$^{-1}$ but are consistent with those from \citet{erb2} who find a median value of $\sim30M_\odot$yr$^{-1}$, although we do not reproduce some of their extreme cases (with SFRs $\sim300M_\odot$yr$^{-1}$). \newline

While a (small) fraction of the disagreement can be explained by a small difference in the adopted upper mass limit for the IMF between
our model and the value adopted by the observational works,
we believe that this is a general problem of semi-analytic models \citep[see, for example,][and references therein]{khoch}. For instance,
if the gas accretion and cooling are uneven over a timestep, the SFR may present spikes that are in better agreement
with the values reported by both \citet{amaze} and \citet{erb06}.
The dynamics of these high redshift galaxies are not always consistent with a smooth star forming
disc, instead they are highly turbulent \citep[see, for example,][]{genz} and a large fraction
do not seem to rotate \citep[for example, those observed by][]{Law}. A similar discussion of the dynamics of the AMAZE galaxies is given in \citet{gner}. This star forming mode is not
yet accounted for in \textsc{G}al\textsc{ICS} and, perhaps, the high value for $\beta$ that we obtain
should be interpreted as a warning: the treatment of the star formation must be improved,
possibly taking into account gas supplies that are enhanced by cold streams.
Namely the required boosting in the SFR is not only given by an increase in the efficiency,
but also by augmenting the gas supply. This has to preferentially happen
in more massive galaxies, where the disagreement with observations occur and, as can be inferred
from the results presented in the previous sections, a much better fit could be obtained by differentially increasing $\alpha$
and $\beta$ at the high mass end with respect to their fiducial values, namely those that correctly fit the
low mass end. On the other hand \citet{Mann} have observed the SFR as a function of $\log\left(M_*/M_\odot\right)$ for another sample of galaxies at $z\sim3$
and have found, on average, a lower SFR than in the AMAZE sample, while at the same time the O abundances are similar to those
we are using in this work. In particular, in the mass range $9.5\le\log\left(M_*/M_\odot\right)\le10.5$ the average SFR that we predict is $\langle \textrm{SFR}\rangle = 30.0 M_\odot$yr$^{-1}$ at $z=2.27$. We note that this value has been corrected to account for the difference in IMFs. The range of the deviation from this average is (approximately) $75M_\odot$yr$^{-1}$ at the three standard deviation level. At $z=3.54$ and for similar masses we find $\langle \textrm{SFR} \rangle = 9.4 \pm 26.4 M_\odot$yr$^{-1}$ at the three standard deviation level.\newline\newline We note that there are many more predicted galaxies in our bin than \citet{Mann} and that the standard deviation in our SFR is due to this large number of predicted galaxies whereas theirs is due to experimental uncertainties involved with their observation. Therefore, another possibility, is that, while AMAZE probes
the systems with the highest SFR, the model adopted here gives values that are more ``typical'' for high redshift
galaxies\footnote{The reader should note that, given all these issues, we chose not to apply
any cutoff based on the SFR to our prediction. Instead we show all the star forming
galaxies in each bin.}. Nevertheless, a successful model should predict also the extreme galaxies, unless those with extreme
SFRs cannot be captured by either the spatial resolution or the lack of physics.
Finally, we note that the observed SFRs discussed above have been derived by assuming
exponentially declining SF histories. \citet{mara} have shown that this assumption
might lead to an overestimate of the SFR with respect to other, more realistic SF histories,
especially when the age is not constrained to be larger then the e-folding time of the assumed
exponentially declining SFR. Similarly, \citet{amaze} have noted that the fits that yield the
SFR in the AMAZE sample sometimes return unrealistically small ages (below 50 Myr) if not suitably constrained.\newline\newline In Fig. \ref{fig:srf} we plot the predicted SFR against stellar mass. It is evident from this Fig. (as well as Fig. \ref{fig:best} discussed below) that galaxies with larger stellar mass and metallicities have higher SFRs and thus our predicted relation is one between SFR and mass, consistent with the recent predictions of \citet{calura} as well as the observational findings of \citet{Mann}. It is also clear from the plots that we do find that, at any given mass, more pristine galaxies have a large SFR and viceversa, in agreement with the recent suggestion by \citet{mann_new} that the mass-metallicity relation is, in fact, a mass-SFR-metallicity surface. However, if we use the new ``variable'' suggested by \citet{mann_new}, namely $\log\left(M_*\right)-0.32\log\left(\textrm{SFR}\right)$ instead of $\log(M_*)$, as the abscissa then we do not find a significant decrease in the scatter. Here we do not make any inferences or draw any conclusions about the relationship that \citet{mann_new} have proposed but leave a more comprehensive comparison for future work where we intend to use a forthcoming, updated and refined version of \textsc{G}al\textsc{ICS} that is better calibrated on the $z=2.27$ and $z=3.54$ mass-metallicity relation.  \newline

\begin{figure}
  \begin{center}
      \subfigure[$z=2.27$\label{fig:sfrz2}]{\includegraphics[width=8 cm,height=8 cm]{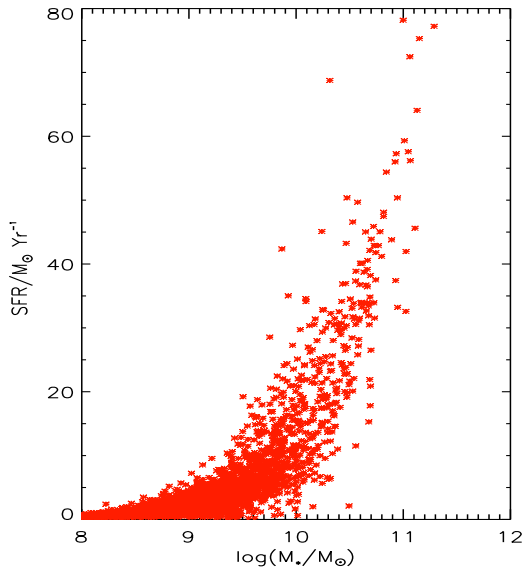}}
      \subfigure[$z=3.54$\label{fig:sfrz3}]{\includegraphics[width=8 cm,height=8 cm]{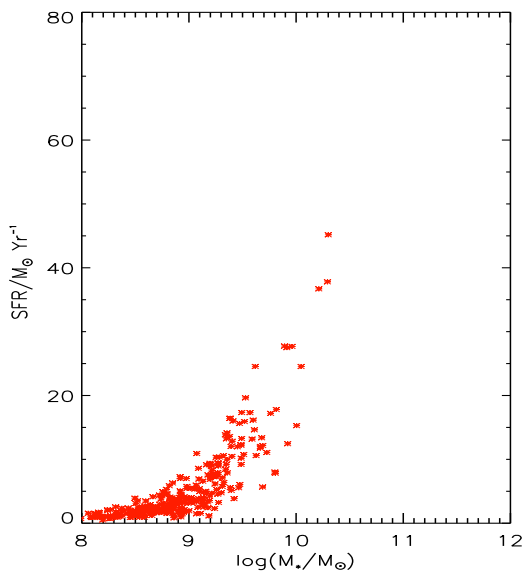}}
      \caption{The predicted SFR as a function of stellar mass for galaxies at redshift $z=2.27$ and $z=3.54$. The red asterisks show individual predicted galaxies with a given stellar mass and SFR.}\label{fig:srf}
\end{center}\end{figure}

%\footnote{Probably not
%to the same degree of local spiral galaxies, in which typical variations along the radius of the gas
%of amount to a 0.3-0.5 dex difference between the centre and
%the edge \citep[see, for example,][]{mag}, the centre being more metal rich.
%Similar gradients are found in the stars of present day elliptical galaxies
%\citep[e.g.][]{oga} and may be expected in their progenitors.}

Our inability to fit the relation at $z=2.27$ may be due to
differences between the regions of the galaxies sampled by observation and by
\textsc{G}al\textsc{ICS}. For instance, the observations of \citet{erb06} are obtained
using a long slit spectrometer with an aperture of $0.75''$, corresponding to a
radius (half-aperture) on the galaxy of only $3.6$ kpc at $z=2.27$ \citep{amaze}.
This implies that \citet{erb06} only sample the central region of the most
massive - hence larger - galaxies, whereas less massive systems
are probed out to larger radii. Local galaxies have metallicity
gradients that, combined with the fixed aperture, bias
the mass-metallicity relation. As shown by Cresci et al. (2010, in press)
high-redshift star forming systems have a complex metallicity structure, however further study is needed to assess its impact on the observed relation. On the theoretical side,
\textsc{G}al\textsc{ICS} calculates the average metallicity over the entire galaxy \citep{galics2},
including the outer regions which are metal poor. We will come back on this
aspect of this problem later when discussing the role of the bulge and the burst components.
AMAZE \citep{amaze} use integral field spectroscopy to measure their metallicities,
extracting the spectral information using a $0.7''$ aperture which corresponds to a
similar radius to the \citet{erb06} sample. Moreover, the observed size evolution \citep[see, for example,][]{law2} of LBGs between $z\sim3.5$ and $z\sim2$,
is rather mild, therefore we deem it unlikely that \citet{erb06} sample a much smaller portion of
these galaxies - at a given fixed mass - than AMAZE, in order to justify
an even larger departure of the models from the observations at $z=2.27$ than at $z=3.54$.
Whilst these effects may be present (indeed we cannot rule them out) and can partly explain the difference between our models' predictions and observations, we
believe that the disagreement at the high mass end has more to do with the treatment of the
physics and the assumptions in the model. \newline

To better explain the lack of chemical evolution in the most massive galaxies we have performed several tests using a chemical evolution model \citep{PM04}
with the same setup (in terms of both IMF and stellar yields) and allowing for different infall and star formation timescales. These models, that follow the evolution of a single galaxy, have the advantage
of giving a self-consistent description of the chemical evolution, whilst using a
simpler parametrisation for the accretion and the star formation histories than we use in \textsc{G}al\textsc{ICS}.
In passing, we note that these models are, in essence, the same ones that are used by \citet{calura} to study the AMAZE galaxies
in the context of their monolithic collapse.
As far as the adopted star formation rate is concerned, it is given by the same formula as \textsc{G}al\textsc{ICS} (equation \ref{eq:SFR}), with the absence of any redshift dependence i.e. $\beta=0$. The star formation timescale $t_{\textrm{SF}}=t_{\textrm{dyn}}/\alpha$ and we assume an exponentially decreasing gas infall law (i.e. $\propto e^{-t/\tau}$, where
$\tau$ is the infall timescale), that is not linked to any Dark Matter accretion history. By running these chemical evolution models for a large
range of different combinations of $\tau$ and $t_{\textrm{SF}}$, we confirm that
a relatively high O abundance ($12+\log\left(O/H\right)>7.5$) is attained within a few 100 Myr.
At the same the time stellar mass is quickly built up. This is why
\textsc{G}al\textsc{ICS}, whose predicted galaxies feature a differing SFR and different accretion/merging histories, predicts a steep mass-metallicity
relation at low masses/younger (relatively) ages.
What differs is the subsequent evolution.
For a star formation timescale much shorter
than the infall timescale, the chemical evolution model show a metallicity $12+\log\left(O/H\right)>8.5$ already at 200 Myr and a very mild evolution after that.
On the other hand, when the star formation timescale is comparable to the infall timescale and both are of the order of 1 Gyr,
galaxies attain a metallicity $12+\log\left(O/H\right)\approx 8$ in the same timescale and a further
0.5 dex increase in the O abundance would take more than 1 Gyr (i.e. the time elapsed between z=3 and z=2). In other words, since this latter case is typical in \textsc{G}al\textsc{ICS},
it is easy for the model galaxies to quickly attain $12+\log\left(O/H\right)\approx 7.5 - 8$ at redshift 3, hence
reproducing the average observed value, but then the predicted metal enrichment in the redshift
range 3-2 is slower than that inferred from the observations because the star formation is
not efficient enough. The former case (star formation timescale shorter than infall timescale), instead, characterises
the \emph{monolithic collapse} models used by \citet{calura} to reproduce the AMAZE data.
It must also be remembered that the $\beta$-boosting at z=3 is by a factor
of 4, whereas the boosting at z=2 is by a factor of 3. Hence, the redshift dependence of the SFR
in such a ``narrow'' range is quite mild and will not necessarily drive major changes in the predicted
mass and metallicities (see Section 4) in this time interval.\newline

Moreover, at higher masses, a fair fraction (approximately 20\%) of model galaxies have had at least one merger.
Therefore some older galaxies could have become quiescent, or at least
have a sizable number of stars in the bulge component.
In this latter case, even if the O abundances are quite
high ($12+\log\left(O/H\right)>8.5$) in the gas associated with the bulge component,
the SFR and the gas mass in this component are much lower than in the disc. Were such a strict
separation between disc and bulge in the models
present also in actual galaxies, the bulge metallicity would not be easily observed.
This seems to be at variance with actual observations \citep[see, for example,][]{law2} that do not detect a drop in the rest-frame UV flux in the central regions of the galaxies. This fact may hint
that the disc/burst/bulge decomposition of model galaxies that well characterises local galaxies
is not a good approximation of the complex structure of high redshift star forming systems. Modifications
such that, for instance, the metal rich gas from the bulge component could enrich the discs, may help
in reconciling such models with observations.\newline

At the same time, as it can clearly be perceived from the fact that there are many more galaxies in the z=2 figures as opposed to the z=3 ones,
we predict a large number of new galaxies that appear during the elapsed time interval. The youngest ones
have stellar masses below $10^{10} M_{\odot}$ and low metallicities, thus biasing the predicted
mass-metallicity relation to lower values and affecting the predicted evolution.
Therefore, in order to improve the agreement with observations, one can apply a SFR/age cut to the
predicted sample of galaxies. Given the difficulties in predicting the correct SFRs - see above -
this has not been attempted.
\newline\newline
A closer inspection of our galaxies show that the metallicity increases with decreasing gas mass
fraction almost as expected in a simple closed box model of chemical evolution, namely $Z\propto ln(M_{tot}/M_{gas})$.
The gas fraction is almost constant (around 0.8) with total baryonic mass (at a fixed redshift) in the high mass
regime, whereas it exhibits a large scatter at the low-mass end. On the other hand it strongly
decreases with stellar mass, although with increasing scatter. This is because the SFR
is constant for galaxies with stellar mass below $10^{10} M_{\odot}$, becoming
proportional to the stellar mass above this limit (c.f. Fig. 9 in \cite{Mann}).
Only for the most massive galaxies (in terms of their stellar mass), does the gas fraction rise again
to (approximately) $0.8$. This is basically what we see in the mass-metallicity plots: the metallicity increases with stellar mass up to $\sim 10^{10} M_{\odot}$
where it then flattens out in the most massive galaxies (where the gas fraction is higher again).
As explained above, in these systems the star formation is not efficient enough compared to the
inflow of primordial gas. Indeed, as shown by \citet{calura}, the high mass end of the z=3 mass-metallicity relation
cannot be reproduced by the ``local'' disc, but proto-spheroids with very high SFR, such that their O
abundance quickly jumps to solar values in a few Myr, are needed.
At variance with such a model, where the galaxy morphology is assumed, \textsc{G}al\textsc{ICS} galaxies may have three
components (disc, bulge and burst, see Section 2 and \citet{galics1}, for more details) that co-exist at the same time
and whose presence is linked to the evolutionary path of the galaxies.
Here we remind the reader that we show the abundances predicted for the disc component of each galaxy only in order to have a fair
comparison with the observations that sample quite a large region compared to the assumed
sizes of both the bulge and \emph{burst} components in the model (whenever they are present). The results do not change if we show the abundances averaged over the three components
because the mass in the cold phase of the disc is much larger than that in the bulge (where we predict
gas mass fraction well below 0.1) or in the burst. In these latter components we predict $12+\log\left(O/H\right)>8.5$ in the majority of the galaxies,
however, such a high O abundance is diluted in the mass averaging. We have already noted that observations sample the inner regions of the galaxies.
On the other hand, we present the results pertaining to the entire disc because the nominal
spatial resolution of \textsc{G}al\textsc{ICS} is $\sim$30 kpc. If we had
enough spatial resolution to make predictions about the inner $\sim$4 kpc only, it is likely that: i)
the \emph{local} density would make the \emph{local} SFR higher, hence the metal
enrichment quicker and ii) the \emph{dilution} of the metals in the bulge and in the burst
by means of the inner disc gas would be lower (meaning a higher metallicity in that region of the disc,
lower fractional contribution in the mass averaging).
Finally, we note that on the abscissa of all the Figs. discussed so far, we plot the total
stellar mass, summed over the three component. Since bulges and bursts tend to be more
frequent at the high mass end, where galaxies are older and had more time to evolve and merge,
this implies that we \emph{stretch} the mass axis without an increase in the O abundance,
hence obtaining an artificially flatter (by a small amount) and more extended to high masses mass-metallicity
relation than the one expected for a pure disc.\newline\newline
One possible way out would involve making predictions that take into account the three component, without
computing an average O abundance. For instance, one could use the discs to explain
the low mass end of the mass metallicity relation, whereas bursts and bulges could explain the high mass end
\citep[see also][]{calura}.
As noted above, the bulges cannot be directly compared with the observations because
it is assumed that the SFR in this component can involve only gas returned from stars,
hence the predicted SFR are lower than those in the disc. Using the bulges instead
of the discs at the high mass end, where they are more abundant, will lead to an increase
in the slope of the predicted mass-metallicity relation, at the expense of an even poorer
agreement between the observed and predicted SFR.\newline\newline
At z=2, nearly 15\% of the galaxies feature a burst component. In such a component, the galaxies exhibit SFRs comparable to, or even a factor of $\sim$2 higher than, the disc
and the predicted O abundances are $8<12+\log\left(O/H\right)<9$, so one may be tempted to use only the
predicted properties of the \emph{burst} only to compare with observations.
The problems with this last scenario are numerous. In the first place, the burst component
is assumed to have a radius that is typically below 1 kpc, hence smaller than
both that of the disc and the aperture used by observers.
We find it difficult to explain the properties of observed galaxies only by means
of such a centrally concentrated component.
Moreover, one has to devise a reason to bias the observations in favour of a ``burst'' \emph{only above} a given mass ($\sim 10^{10}M_{\odot}$) in order to steepen the mass-metallicity relation at the high mass end. In part, such a bias can be
granted by the fact that observed samples are selected according to their SFR\footnote{At least at z=2. In fact,
the predicted SFRs in the burst components at z=3 are still lower than the observed ones.}. However, since
the model switches the burst component on as a consequence of mergers, the hypothetical solution of having a greater frequency of burst components
in the more massive galaxies is at variance with the fact that a fraction of the observed galaxies do not show such signatures.
Moreover, recent observations \citep[for example those of][]{forster} indicate that galaxies that are observed with a regularly rotating disc are preferentially
higher mass systems. However, these discs still have large amounts of random motions and turbulence can be fairly thick.
Future developments of the models should implement the possibility that ``bursts'' are not
only centrally located. Instead SF clumps would be distributed in the disc, as also discussed above
when comparing predicted and observed SFRs.
\newline

The net effect of the changes suggested above should create a typical galaxy formation path in which, broadly speaking,
$\alpha$ increases with galactic mass. Indeed a direct increase of $\alpha$ with mass may be achieved
by directly incorporating the suggestion by \citep{pipinosfr} on the positive feedback by the central black hole.
Such a change can be easily implemented in monolithic formation models, however it is more difficult to devise
a mechanism in a scenario where larger galaxies are formed through mergers of smaller companion.
This is where the study of the mass-metallicity relation at $z\approx3$ provides leverage:
many (predicted) galaxies have not undergone any mergers yet and sot he effect of these mergers is kept at a minimum level. These mergers tend to produce high mass galaxies with lower metallicities than those that have evolved in the standard manner and thus (in general) flatten the relation. Hence the observed mass-metallicity relation is, in fact, giving us insight into the SFR-mass and inflow-mass relations. To be more quantitative, from the investigation done in Section 4 we can infer that
if the value of $\alpha$ is kept at the fiducial \textsc{G}al\textsc{ICS} value at the low mass end, whereas its value
is increased by a factor of at least 3 at the high mass end, the predicted mass-metallicity relation would be considerably steeper
and in better agreement with observations. The reader should note that this is an illustrative example to provide insight into the manner in which current models should be modified in order to improve them in this respect and that unnaturally fixing $\alpha$ as a function of stellar mass is only an artificial solution and is not the correct method by which this can be achieved. In passing, we note that these solutions may also alleviate the problems that semi-analytic models have in reproducing the abundance ratio-mass
relations in present-day elliptical galaxies \citep{galics2}.\newline

\begin{figure}
  \begin{center}
      \subfigure[$z=2.27$\label{fig:asala}]{\includegraphics[width=8.1 cm,height=8 cm]{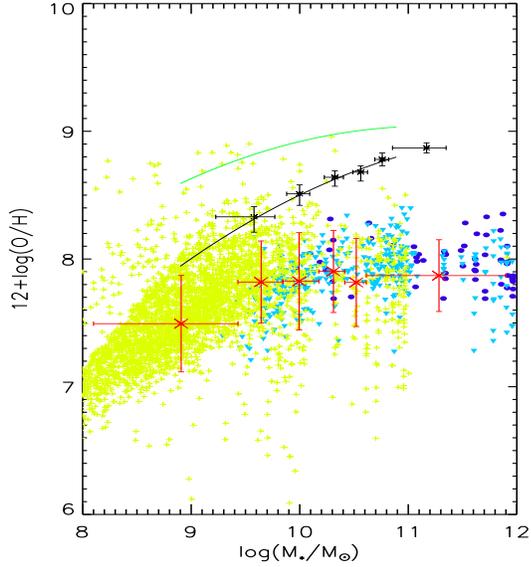}}
      \subfigure[$z=3.54$\label{fig:asalb}]{\includegraphics[width=8.1 cm,height=8 cm]{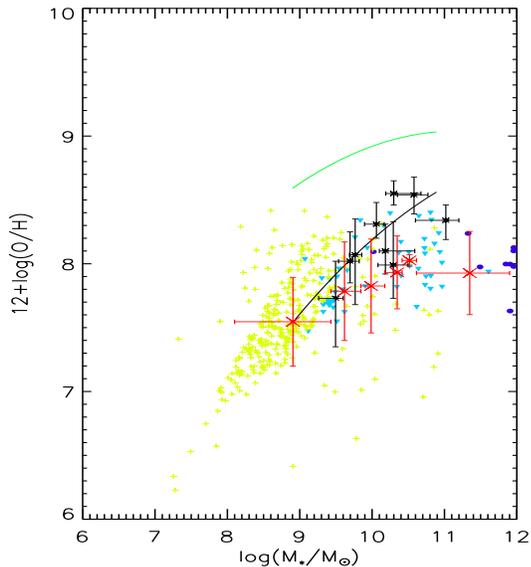}}
      \caption{Predicted mass-metallicity relation using the best-fitting parameters $\alpha=0.02$, $\beta=1.0$, $\varepsilon=0.15$ and $\zeta=0.3$ and the Salpeter IMF (equation \ref{eq:ken}).
      The yellow crosses represent galaxies with $\textrm{SFR}<10 M_\odot \textrm{yr}^{-1}$, the light blue triangles represent galaxies with $10\le\textrm{SFR}<30 M_\odot \textrm{yr}^{-1}$
      and the dark blue circles represent galaxies with $\textrm{SFR}\ge30 M_\odot \textrm{yr}^{-1}$. The green (upper) line shows the AMAZE best-fitting line for $z=0.07$ and the black asterisks show the data from
      \citet{erb06} (\ref{fig:asala}) or \citet{amaze} (\ref{fig:asalb}). The black (lower) line shows the AMAZE best-fitting line for $z=2.27$ (\ref{fig:asala}) or $z=3.54$ (\ref{fig:asalb}). The red crosses show
      the average mass and metallicity in each of the 6 mass bins used by \citet{erb06}, the horizontal bars show the mass bins and the vertical bars show one standard deviation in the metallicity.}\label{fig:best}
  \end{center}
\end{figure}

We also note that, although variation of the parameters significantly
changes the prediction of the mass-metallicity relation at high
redshift, the effect at low redshift is much less evident.\newline

\begin{figure}
  \begin{center}
      {\includegraphics[width=8.1 cm,height=8 cm]{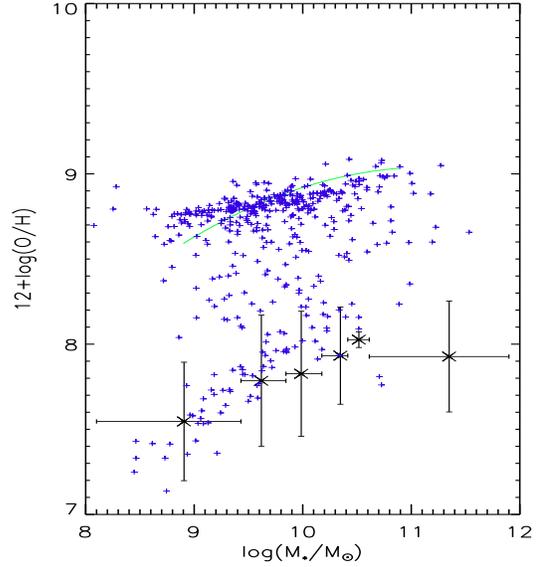}}
      \caption{The predicted evolution from $z=3.54$ to $z\approx0$. The black crosses show the average mass and metallicity in each of the six mass bins used by \citet{amaze} at $z=3.54$ using the best-fitting parameters.
       The horizontal bars show the range of each mass bin and the vertical bars show one standard deviation in the metallicity. The AMAZE
       best-fitting curve (equation \ref{eq:cal}) at $z=0.07$ is shown by the green line and the blue crosses show the predicted mass-metallicity relation at $z\approx 0$.}\label{fig:z=0}
  \end{center}
\end{figure}

On the other hand, examination of the
predicted distribution at $z=0$ shows that the evolution in
metallicity between $z=3.54$ and $z=0.07$ agrees with the observed
evolution \citep{trem04,amaze}. This is shown in Fig.
\ref{fig:z=0}. Previous models (for example \citet{del04}, \citet{koba} and others discussed in the introduction) often fit the relation at one redshift only and the redshift evolution is not tested to $z=0$ (see \citet{amaze} section 7.5 for a discussion). We have fitted the observed relations at both $z=3.54$ and $z\approx0$ and hence the model is more tightly constrained. We have not been entirely successful in fitting the observed relation at all redshifts however we have attempted to constrain it in different regimes and predict the observed evolution of the relation from $z=3.54$ to $z=0$. Furthermore, we have found that outflows are not the origin of the relation and only effect its low redshift evolution and hence we have not needed to impose different outflow models to fit at specific redshifts. Nor have we needed to fine-tune the efficiency of supernova feedback.

\section{Conclusions}
\label{sec:conc}

The relation between stellar mass and gas-phase metallicity has been
well documented over the range $0\le z\le 3.54$
\citep{trem04,erb06,amaze}. Despite this many hydrodynamical and
semi-analytic $N$-body simulations of galaxy formation have been
unable to predict the correct relation over a suitable redshift range
\citep{del04,derossi,Mouchine}. In this paper we have
fitted the observed relation at $z\approx 3.54$  using an updated version
of {\sc G}al{\sc ICS} given in \citet{galics2} and have reproduced the observed relation at $z\approx0$. The model uses a $\Lambda$CDM
cosmology taken from WMAP3 \citep{wmap3} and includes outflows due
to type-II supernovae since previous work \citep{derossi} has shown
that these are needed to reproduce the observed relation. Many free
parameters are included that control the star formation efficiency,
redshift dependence of the SFR, type-II supernova feedback
efficiency and the
halo recycling efficiency which were varied separately in order to achieve the best possible fit to observation \citep{amaze}.
Here we summarise the results of the simulations.\newline\newline
Firstly, we have found that the effect of varying the supernova feedback efficiency has little impact on the relation at high redshift but does have a small effect at lower values. We have also predicted that more massive galaxies have higher SFRs than less massive ones. These two results taken together support recent findings \citep{Mann,calura} that it is the SFR-mass relation in galaxies and not galactic outflows that is responsible for the origin and evolution of the relation although the cumulative effects of outflows do affect the low redshift evolution. Secondly, a better agreement between the predicted and the observed average metallicity at a given redshift can be obtained by assuming that the SFR that
has a strong redshift dependence, proportional to $1+z$, which is slightly stronger than other models have used in the past (\citet{cat2} uses $\left(1+z\right)^{0.6}$). At the same time, the predicted SFRs are lower than the observed values at $z\sim3$, but are consistent at lower redshifts. These facts
may point to the need for a revision of the SFR recipes in a future generation of semi-analytical models.
Thirdly, the observed relation at $z=3.54$ is reproduced well with a
scatter in the distribution of stellar mass and metallicity
comparable to observations \citep{amaze}. However, the predicted relation is flatter at the high mass end.
Also, we fail to reproduce the relation at $z=2.27$ \citep{erb06}, where the flattening in the
high mass regime becomes more evident. We discuss several reasons for this disagreement, stemming from both theoretical and
observational biases. We argue that if observations are preferentially
selecting galaxies with high SFRs and the measured abundances mirror those in these
``bursts'' rather than averages over the three components, we might solve the problem of the flattening by
using the abundances predicted in the ``burst'' rather than in the disc component of the
model galaxies. This is a rather ad hoc solution and presumably means that
future models should take into account that, at high redshifts, either the SF occurs in clumps
rather than in a smooth disc or the star formation efficiency is boosted
preferentially in the most massive galaxies by, for example, positive AGN feedback \citet{pipinosfr}. We note, that the model does reproduce the observed evolution
to $z=0$. The efficiency of
outflows was found to have only a minimal effect on the predicted distribution. Finally, we have seen that
low-mass galaxies are best fitted to the relation using a low star
formation efficiency whereas the converse is true regarding
high-mass galaxies. This observation supports the findings of
\citet{amaze} and thus galaxy downsizing may be the origin of the ubiquitously observed
relation.

\section*{Acknowledgments}
We graciously thank the anonymous referee for their comments and suggestions, all of which
have greatly improved this paper. This work was partially supported by the Italian Space Agency
through contract ASI-INAF I/016/07/0.

\bibliography{origin}
\bibliographystyle{mn2e}
\end{document}